\documentclass[aps,prf]{revtex4}{

\usepackage{graphicx}

\usepackage{amsmath,amsfonts,bm, color,amssymb,float}
\usepackage{graphicx,epsfig,overpic,hyperref,ulem,caption,subcaption}%% OPTIONAL MACRO DEFINITIONS

\newcommand{\out}{{\mathrm{out}}}
\newcommand{\ins}{{\mathrm{ins}}}

\newcommand{\surf}{{\mathrm{s}}}

\newcommand{\FJ}{FJ}

\newcommand{\Ca}{\mbox{\it Ca}}
\newcommand{\bCa}{\mbox{\textbf{\textit{ Ca}}}}%bold Ca
\newcommand{\Ma}{\mbox{\it Ma}}

\newcommand{\bomega}{{\bm \omega}}

\newcommand{\bR}{{\bf \hat d}}
\newcommand{\bv}{{\bm{u}}}
\newcommand{\bV}{{\bm{U}}}
\newcommand{\bX}{{\bm{X}}}

\newcommand{\bu}{{\bm{u}}}

\newcommand{\Rdrop}{R_0}

\newcommand{\xhat}{{\bf \hat x}}
\newcommand{\yhat}{{\bf \hat y}}

\newcommand{\rhat}{{\bf \hat r}}
\newcommand{\phat}{{\bf \hat p}}
\newcommand{\that}{{\bf \hat t}}

\newcommand{\nhat}{{\bf \hat n}}
\newcommand{\bx}{{\bf x}}
\newcommand{\bxz}{{\bf x_0}}
\newcommand{\by}{{\bf y}}

\newcommand{\drop}{d}
\newcommand{\particle}{p}
\newcommand{\charstrength}{Q}
\newcommand{\bnab}{{\bm{\nabla}}}
\newcommand{\st}{\gamma}

\newcommand{\undisplaced}{{\mathrm{act}}}
\newcommand{\displaced}{{\mathrm{act}}}
\newcommand{\visc}{\mu}
\newcommand{\viscratio}{\lambda}
\newcommand{\charforce}{F}
\newcommand{\ysp}{y}
\newcommand{\Yspp}{Y}
\newcommand{\surfactant}{surfactant}
\newcommand{\q}{\sigma}
\newcommand{\bt}{{\bf{t}}}
\newcommand{\bT}{{\bf{T}}}
\newcommand{\Sp}{S^+}
\newcommand{\Tp}{S^-}
\newcommand{\sumjmq}{\sum_{j,m,\q}}

\newcommand{\steadyshape}{\mathrm{ss}}

\newcommand{\refeq}[1]{Eq. (\ref{#1})}
\newcommand{\reffig}[1]{Fig. (\ref{#1})}
\newcommand{\refsec}[1]{Section (\ref{#1})}
\newcommand{\refapp}[1]{Appx. (\ref{#1})}

\begin{document}

%\title{Dynamics of a deformable droplet enclosing an active particle}
\title{Migration and deformation of a droplet enclosing an active particle}
\author{Sho Kawakami and  Petia M. Vlahovska}
% \corresp{\email{petia.vlahovska@northwestern.edu}}}

\affiliation{Engineering Sciences and Applied Mathematics, Northwestern University, Evanston, IL 60208, USA}

\begin{abstract}
The encapsulation of active particles, such as bacteria or active colloids, inside a droplet gives rise to nontrivial shape dynamics and droplet motility. To understand this behavior, we derive an asymptotic solution for the fluid flow about a deformable droplet containing an active particle, modeled as a Stokes-flow singularity, in the case of small shape distortions. Offsetting of the active particle from the center of the drop breaks symmetry and leads to excitation of large number of shape modes as well as particle and drop displacement. Flows due to common singularity representations of active particles, such as Stokeslets, rotlets, and stresslets, are computed and compared to results for non-deformable droplets enclosing active particles. The effect of interfacial properties is also investigated.  Surfactants adsorbed at the droplet interface  immobilize the interface and arrest the droplet motion. Our results highlight strategies to steer the flows of active particles and create autonomously navigating containers.

\end{abstract}

\date{ \today}

\maketitle

\section{Introduction}

Many biological cells are capable of autonomous locomotion.  
Bacteria, for example, exhibit directed motion as they sense and move towards nutrients while navigating complex environments \citep{Bastos-Arrieta:2018}. 
Artificial systems that mimic this type of behavior possess great  potential 
for the engineering of autonomous micro-robots \citep{li2017micro,Lee-Shields:2023}, however achieving internally driven motility is a challenging task.  Active matter, which consists of entities (active particles) capable of harvesting
energy from the environment and converting it into motion, presents a promising solution of this problem.  The constant energy dissipation unlocks a wealth of phenomena in active matter that are impossible at equilibrium, e.g.,  self-organization and directed coherent motion on scales much larger than the individual particle \citep{Marchetti:2013}.
%steered driven motility  
%driven internally by its constituents, which transform stored free energy into motion
Bacteria and swimming microorganisms are  examples of living active particles \citep{Bechinger16,Elgeti15,Gompper:editorial2021}. There has been a great effort to create  self-propelled particles 
emulating microswimmers \citep{SHIELDS:2017,Ebbens:2018,BBBreview:2023,Bharti:2022,Zarzar:2021,BOYMELGREEN:2022,Diwakar:2022,Michelin:2023}. However, 
%while there has been progress in emulating the single particle dynamics like run-and-tumble dynamics \cite{Karani:2019}, 
effective strategies to control the collective dynamics of such artificial micromotors without external steering,  or utilize micromotors to move cargo-carrying containers  remain elusive.

Recently, 
spontaneous displacement of a cell-like soft container enclosing active particles
has been achieved experimentally using a droplet containing motile colloids \citep{kokot2022spontaneous} or bacteria \citep{Ramos:2020,Rajabi:2021}. 
While one can intuitively appreciate that the activity of the particles drives droplet motion, a comprehensive understanding of the mechanisms underlying droplet self-propulsion is necessary to design a strategy to effectively control the droplet locomotion. 
%Control over direction has been accomplished by using strutured environment such as liquid crystals \citep{Rajabi:2021}.
 It is well known that geometric boundaries strongly influence particle dynamics and often are used to orchestrate the collective behavior of active ensembles.
For example, while unconfined suspensions of bacteria exhibit turbulent-like flow \citep{Wensink:2012, Alert:2022}, directed motion emerges when the  suspension is confined to a channel \citep{Wioland:2016} or  a macroscale  vortex forms when the bacteria are constrained in a droplet \citep{Wioland:2013,lushi2014fluid}.
%or self-organize into lattices of hydrodynamically bound vortices with a long-range antiferromagnetic order when constrained by two-dimensional arrays of microscopic pillars~\cite{nishiguchi2018engineering}.  
Similar behavior is observed with 
%formation of a macroscale vortex is observed with 
%coherent vortical motion emerges in collectives of  
Quincke rollers constrained by solid boundaries \citep{bricard2013emergence,bricard2015emergent,Chardac:2021}.
%formation of a macroscale vortex is observed with the Quincke rollers \cite{Bartolo:2015} under confinement.

In the case of the droplet, the boundary is soft and can deform in response to the internal flow generated by the active particles. The flow about a self-propelled active particle such as microswimmers or active droplets in unbounded fluid is well known \citep{Lauga:2016,Saintillan:2018,Lauga_2020,Michelin:2023,Ishikawa:2024}. However, active particles in a spherical container  are less studied  and most work is focused on modeling particle motion in a rigid enclosure \citep{zia:2016,zia:2018,chamolly2020stokes,marshall2021hydrodynamics}. 

 In the case of a non-deformable spherical drop, exact solutions for a microswimmer modeled as a squirmer at the center the droplet along with boundary integral simulations of the squirmer placed in an arbitrary location inside the drop were first considered in \cite{reigh2017swimming,reighPRF}. Subsequent  works developed analytical solutions  also for non-axisymmetric configurations \citep{kree2021controlled,kree2022mobilities}, or surfactant-covered drop \citep{shaik2018locomotion}.
 %The droplet  motility due to many enclosed squirmers has been analyzed in \cite{huang2020active}. 
 Models of the active particle as a point singularity have examined the case of a point force, a stresslet, and a rotlet placed at an arbitrary position inside the drop \citep{daddi2018creeping,hoell2019creeping,sprenger2020towards,kree2021dynamics}. These works reported that  even a force-free singularity can give rise to net droplet translation.
%where a non-zero velocity for drops containing point forces and force dipoles were found.
 Deformation in response to the active internal flow was solved analytically in the case of an elastic shell, assuming the container shape remains close to a sphere \citep{hoell2019creeping}.
 %{\col{[comment. is the solution correct?.] }}
 %The flows due to singularities  placed outside of a drop have also been considered \citep{shaik2017point}.
Large deformations of a container due to active particles have only been considered in the absence of hydrodynamic interactions \citep{paoluzzi2016shape,wang2019shape,quillen2020boids,Urlap:2023, Lee:2023}. In these simulations,  the boundary is modeled as a chain of spring-connected beads  and the container deformation arises solely from the collisions  between active particles and boundary beads. Simulations with full hydrodynamics of many microswimmers are limited to a non-deformable, spherical drop \citep{huang2020active} .

In this work, we  address the case of an active particle, represented by an arbitrary point singularity, inside a slightly deformable surfactant-free and surfactant-covered drop. 
We adapt the methods developed to model deformable drops in an applied external flow \citep{vlahovska2009small,vlahovska2015dynamics} to the flow generated inside the droplet by  a Stokes flow singularity. 
Drop deformation and migration velocity are obtained, and the feedback on  the  trajectory of a swimming active particle is explored. 
The paper is organized as follows.  
\refsec{sec:Formulation} presents the formulation of the problem and \refsec{sec: Solution Method} provides the general solution method. 
\refsec{sec:Results} presents results for the flows generated by the Stokeslet, rotlet and stresslet. 
The shape and velocity of the drop are then calculated. General results for higher order singularities inside the droplet are also given. 
The case of a stresslet inside a droplet is used to illustrate the impact of the transient internal flow on the motion and reorientation of the enclosed active particle.
%{\col{drop shape on the dynamics of an enclosed active particle.: really - the flow at steady state is identical to.a sphere?}}

\section{Problem formulation}
\label{sec:Formulation}

\begin{figure}
\centering
\begin{subfigure}[b]{0.25\textwidth}
\includegraphics[width=\linewidth]{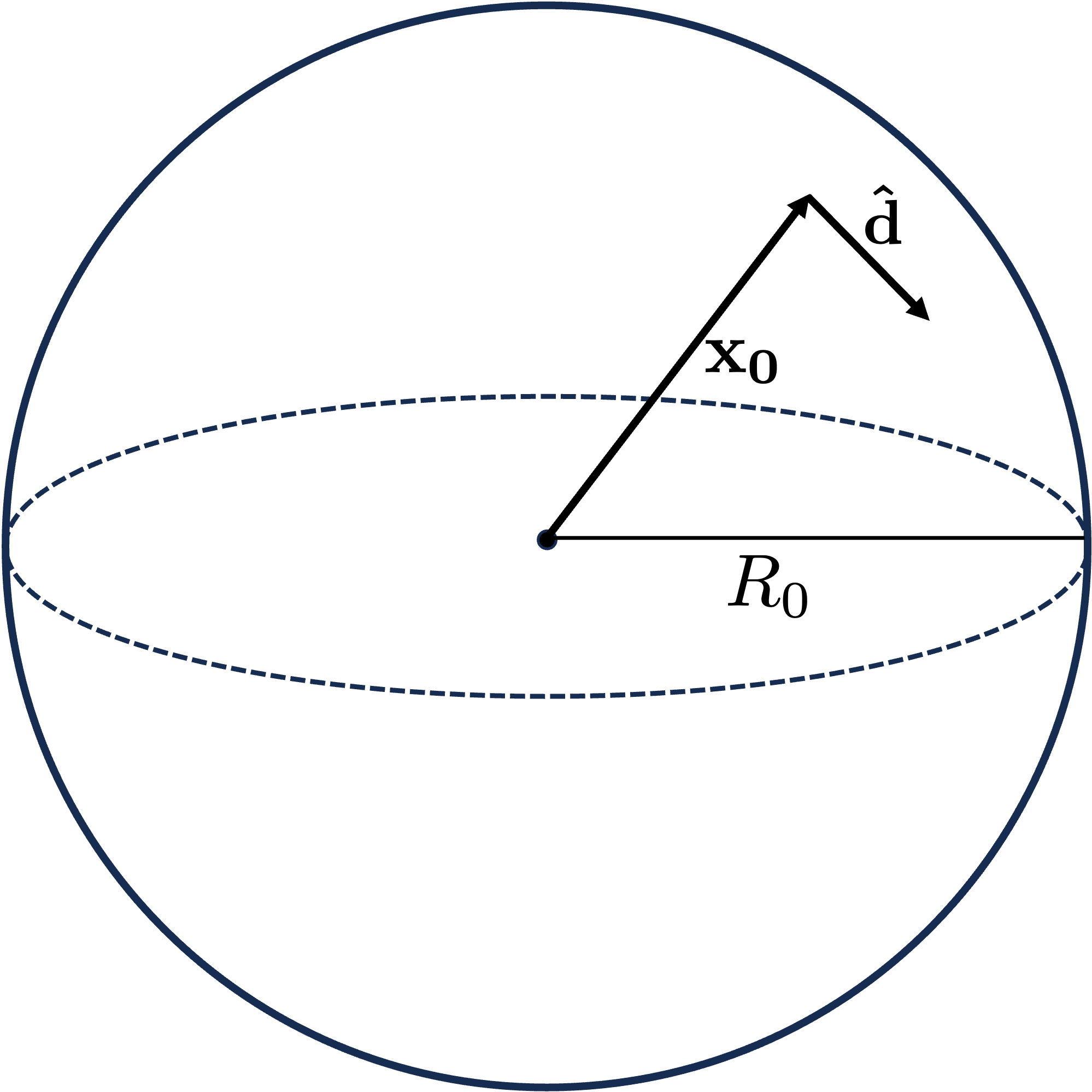}
\caption{}\label{table: singularities1}
\end{subfigure}
\begin{subfigure}[b]{0.7\textwidth}\hfill
{\tiny
\begin{tabular}{|l|l|c|c|}
\hline
Singularity & Solution in unbounded fluid $\bv(\bx)$ & $\bm{s}(\bx)$& $q(\bx)$ \\\hline\vspace{5pt}
Stokeslet & $\frac{1}{8\pi\viscratio |\bx|}(\bR+\frac{(\bR\cdot \bx)\bx}{|\bx|^2})$& $\bR\delta(\bx)$&0\\\vspace{5pt}
Rotlet &$\frac{\bR \times \bx}{8\pi\viscratio |\bx|^3}$&$\frac{1}{2}\bnab \times (\bR\delta(\bx))$&0\\\vspace{5pt}
Axisym. Stresslet &$\frac{1}{8\pi\viscratio |\bx|^2} (-\frac{\bR\cdot\bx}{|\bx|}\bR+3\frac{(\bR\cdot\bx)^2}{|\bx|^3}\bx)$&$\nabla \cdot((\bR\bR-\frac{1}{3}|\bR|I) \delta(\bx))$&0\\
Source Dipole & $\bv = \frac{1}{4\pi|\bx|^3}(-\bR+3\frac{\bR\cdot \bx\bx}{|\bx|^2})$&0 &$\bR\cdot\bnab \delta(\bx-\bxz)$\\
\hline
\end{tabular}}
\caption{}\label{table: singularities2}
\end{subfigure}
%\caption{Table of point force and source densities in Equation \ref{eq: StokesEqinside} and the resulting velocity $\bv(\bx)$ in unbounded domain. See \cite{graham2018microhydrodynamics} for detailed derivations.}\label{table: singularities}
\caption{(a) Sketch of the problem: a Stokes flow singularity ( Stokeslet ) inside of a droplet. (b) The Stokes flow singularities used to model the active particle in our analysis \citep{graham2018microhydrodynamics} .}
\label{table: singularities}
\end{figure}

Consider an initially spherical, neutrally buoyant droplet with equilibrium radius $\Rdrop$ and viscosity $\viscratio \visc$ suspended in an unbounded fluid with viscosity $\visc$. 
The droplet interface is either clean with interfacial tension $\st_0$ or covered with an insoluble non-diffusive surfactant monolayer with equillibrium interfacial tension $\st$.
An active particle with strength $\charstrength$ and swimming velocity $V_\particle \phat$ is placed inside the droplet, see \reffig{table: singularities1} for a sketch of the problem.
We will model the active particle as a Stokes flow singularity, e.g. a Stokeslet, a stresslet, a rotlet, and a source dipole, see Table in \reffig{table: singularities2}.

Hereafter all quantities are rescaled using the droplet radius $\Rdrop$ and the characteristic active force $\charforce$.
In the case of the potential flow singularities, i.e., the source multipoles, 
\begin{equation}
\charforce = \frac{\charstrength \visc\viscratio}{\Rdrop ^{j+1}}
\end{equation}
where $j=1$ corresponds to a source dipole, $j=2$ corresponds to a source quadrupole, and so on.
In the case of the force singularities, 
\begin{equation}
\charforce = \frac{\charstrength }{R^j}
\end{equation}
where $j=0$ corresponds to a point force, $j=1$ to a force dipole, $j=2$ to a force quadrupole and so on.

The droplet deformation due to the flow generated by the active particle is characterized by the capillary number
\begin{equation}
\Ca = \frac{\charforce}{\gamma \Rdrop }.
\end{equation}

For the surfactant-covered drop, the surfactant redistribution gives rise to gradients in the surface tension. Their strength relative to the active flow stresses is quantified by the Marangoni number,
% there is an additional dimensionless parameter, the Marangoni number:
\begin{equation}
\Ma = \frac{(\st-\st_0) \Rdrop }{\charforce},
\end{equation}
%which gives the ratio of the distorting viscous and restoring Marangoni stresses. 
We consider the regime of $\Ma$ large so that the surfactant coverage remains nearly uniform, i.e. $\Ma \Ca  \sim O(1)$.

The fluid velocity and pressure inside the drop, $\bv^{\ins}$ and $p^{\ins}$, and outside the drop, $\bv^{\out}$ and $p^{\out}$, are described by the Stokes equation:
\begin{align}
-\bnab  p^{\out}+\nabla^2 \bv^{\out} = 0 &\, ,\,\,\,\,\, \bnab \cdot \bv^{\out} = 0\label{eq: StokesEqoutside}
\\
-\bnab  p^{\ins}+\viscratio\nabla^2 \bv^{\ins} = \bm{s} &\, ,\,\,\,\,\, \bnab \cdot \bv^{\ins} = q .\label{eq: StokesEqinside}
%-\viscratio\bnab p^{in}+\viscratio\nabla^2 \bv^{in} = \bm{s} &\, ,\,\,\, \bnab \cdot \bv^{in} = \bm{q}
\end{align}
%\end{split}\label{eq: StokesEq}
%\end{equation}
The point distributions $\bm{s}(\bx-\bxz)$ and $q(\bx-\bxz)$ model the active particle located at $\bxz$ in a coordinate system centered at the droplet, see
%contributions of the active particle on the flow.
%Some common models such as a point force, force dipole, rotlet, and source dipole are given in 
the table in \reffig{table: singularities2}.
The position of the droplet interface is specified by $\bx_\surf = r_\surf(\theta,\phi,t)\rhat$. It evolves according to 
\begin{equation}
\frac{d r_\surf}{dt} = \bv_\surf \cdot \nhat ,\label{Eq: surface evolution}
\end{equation}
where $\nhat$ is the outward normal vector and $\bv_\surf$ is velocity of the interface, 
\begin{equation}
\bv^{\ins} = \bv^{\out} = \bv_\surf, \,\, \text{at }r=r_\surf  \label{BC: continuity}.
\end{equation}

%\begin{table}

Fluid motion gives rise to bulk hydrodynamic stress,
\begin{equation}
\bm{T}^{(i)} = -p^{(i)}\bm{I}+\eta^{(i)}\left[\nabla \bv^{(i)}+(\nabla\bv^{(i)})^T\right] \label{Eq: bulkstress}
\end{equation}
where $(i) \in \{\ins, \out\}$ and $\eta^{\out} = 1, \eta^{\ins} = \viscratio$.

For a clean drop, the jump in normal stress at the interface is balanced by the surface tension, and the tangential stresses are continuous across the interface:
\begin{equation}
\nhat \cdot (\bm{T}^{\out}-\bm{T}^{\ins}) = \Ca^{-1} \left(\bnab \cdot\nhat \right)\nhat,\,\,\,r=r_\surf \label{BC: normal stress balance}.
\end{equation}

For a surfactant-covered drop, %{\color{red}Changed to second one since we are using surface tension not surfactant concentration}
\begin{align}
%&\bm{\nabla_s}\cdot \bv_s = 0\label{BC: surface incompressibility surfactant}\\
%&\nhat \cdot (\bm{T}^{\out}-\bm{T}^{\ins}) = \Ca^{-1} \st \left(\bnab \cdot\nhat \right)\nhat+ \Ma \bnab _S \st,\,\,\,r=r_\surf. \label{BC: stress balance surfactantp}\\
&\nhat \cdot (\bm{T}^{\out}-\bm{T}^{\ins}) = \Ca^{-1} \left[\st \left(\bnab \cdot\nhat \right)\nhat+  \bnab _\surf\st\right],\,\,\,r=r_\surf. \label{BC: stress balance surfactant}
\end{align}

The active particle swims and it is advected and reoriented by the fluid flow. Its trajectory is given by
%An active particle enclosed in the droplet is given an orientation $\phat , ||\phat|| = 1$. The evolution equation for the particle location $\bm{X_\particle}$ and orientation $\bm{\hat{p}}$ is given by
\begin{equation}
\begin{split}
&\frac{d \bm{\bxz}}{dt} = \tilde{V}_\particle \phat+\bv_\particle \,\,\quad \frac{d\phat}{dt} = \bomega_\particle \times \phat
\end{split}\label{Eq: Particle Evo Eq}
\end{equation}
where the normalized particle velocity is $\tilde{V}_\particle = V_\particle \visc \Rdrop /\charforce$, and $\bv_\particle, \bomega_\particle$ are the local flow velocity and rotation rate.
%translational and rotational velocity imparted onto the particle by the flow.

\section{Asymptotic solution for small shape deformation}\label{sec: Solution Method}

We solve the problem in the case of weak particle activity, i.e. $\Ca \ll 1$, where the droplet shape remains nearly spherical.
Given the spherical geometry of the problem, we expand all variables in spherical harmonics (see \refapp{Ap:vel fields}).
Accordingly, the surface in a coordinate system centered at the droplet is parametrized as 
%The surface of the drop is assumed to be nearly spherical and subject to small deformations characterized by parameter $\epsilon\ll 1$. Thus the surface is parametrized as
\begin{equation}
r_\surf(\theta,\phi,t) = 1+\epsilon f(\theta,\phi,t).\label{Eq: radial par1}
\end{equation}
where $\epsilon \equiv \Ca$ and the deviation from sphericity is expressed in terms of scalar spherical harmonics 
%At any fixed instant in time $t$, the location of the active particle $\bxz$, as well as the force and source distributions $\bm{s}$ and $q$ will be specified and the flow will be given in terms of asymptotic expansions in $\epsilon$. 
%We will limit this work to linear $O(\epsilon)$ order. 
%The shape deformation $f$ is expressed in terms of scalar spherical harmonics :
\begin{equation}
f(\theta,\phi,t) = \sum_{j=1}^\infty \sum_{m=-j}^{j}f_{jm}(t) \ysp_{jm}(\theta,\phi).
\end{equation}
In the case of a surfactant-covered drop, we also assume that the surface tension remains nearly uniform
%Additionally for the surfactant covered droplet, the tension is also subject to small deviations and are likewise expanded in terms of scalar spherical harmonics
\begin{equation}
\st = \st_0+\epsilon \sum_{j=1}^\infty\sum_{m=-j}^{j} g_{jm}(t)\ysp_{jm}(\theta,\phi),\label{eq: surface incompressibility}
\end{equation}
which implies that the surface flow is area-incompressible at leading order \citep{stone1990simple,blawzdziewicz2000rheology,vlahovska2002nonlinear}
%The surfactant transport equation \citep{stone1990simple} under nearly uniform concentration, reduces to surface flow incompressibility,
\begin{equation}
\bnab_\surf \cdot \bv_\surf = 0.\label{Eq: surface incompressibility}
\end{equation}

\subsection{Solution for the fluid velocity}
The fluid flow inside and outside the droplet are expanded in a basis of fundamental solutions of the Stokes equation in a spherical geometry, $\bv_{jm\q}^\pm(\bx)$, where $\pm$ denotes solutions that are regular at infinity $(-)$ or at the origin $(+)$, and $\q=0,1,2$ denote the irrotational, rotational, and pressure-conjugate velocity fields (see \refapp{appendix: spherical harmonics} for details).

The flow velocity inside and outside the droplet is computed in three steps.

\begin{enumerate}
\item 
%% Part 1 the unbounded solution
The flow due to the active particle in free space filled with the droplet fluid, which is the solution $\bv(\bx,\bxz)$ of \refeq{eq: StokesEqinside} in a coordinate system centered at the singularity, is expressed in decaying solutions of the Stokes equation
%the representation in terms of decaying solutions to the Stokes equation centered at itself (e.g. set $\bxz =\bm{0}$) is derived.
%We will refer to this singularity centered at the origin as the undisplaced singularity:
\begin{equation}
\bv^{\undisplaced}(\bx) = \sumjmq c_{jm\q}^{\undisplaced}\bu_{jm\q}^-(\bx)\,, \label{Eq:undisplacedsingularity}
\end{equation}
where $\sumjmq = \sum_{j=1}^\infty \sum_{m=-j}^j\sum_{\q=0}^2 $.
%The coefficients can be found by using the orthogonality of the Stokes solution basis, $c_{jm\q}^{\undisplaced} = \langle\bv(\bx,\bm{0}),\bu_{jm\q}^-(\bx) \rangle = \int_\Omega \bv(\bx,\bm{0}) (\bu_{jm\q}^-(\bx))^\ast d\Omega$, where $\Omega$ is the surface of a unit sphere.
For example a rotlet with axis of rotation $\bm{\hat{z}}$ is described by coefficient $c_{101}^\undisplaced = -\frac{i}{2\sqrt{6\pi}\viscratio}$.

%%Displacement
\item %The flow that deforms the droplet needs to be computed in a reference frame centered at the droplet.
Droplet deformation is computed from \refeq{Eq: radial par1}. For this purpose, the flow \refeq{Eq:undisplacedsingularity} needs to be transformed  into a coordinate system centered at the droplet.
A similar problem is encountered in the study of electrostatic and gravitational fields, where the field at distant points is sought in terms of sources in a given region, and is solved by the multipole expansion method \citep{jackson1999classical}. 
The difference in our case is that we are dealing with a vector (velocity) not scalar (potential) fields.
 This problem has been solved by \cite{felderhof1989displacement}.\\

The structure of the transformed flow depends on the position $\bx$. In the region $|\bx|<|\bxz|$, the velocity field is a superposition of Stokes flow solutions that are non-singular at the origin. In the region $|\bx|>|\bxz|$, the flow consists of Stokes flow solutions that are non-singular at infinity
\begin{equation}
\bv^{\displaced}(\bx,\bx_0) = \sumjmq c_{jm\q}^{\displaced,+}(\bxz)\bu_{jm\q}^+(\bx)H(|\bx_0|-|\bx|)+c_{jm\q}^{\displaced,-}(\bxz)\,,\bu_{jm\q}^-(\bx)H(|\bx|-|\bx_0|)
\end{equation}
where $H$ is the step function. 
The coefficients of the transformed flow $c_{jm\q}^{\displaced,\pm}$  are determined by the transformations derived in \cite{felderhof1989displacement} and the details are given in \refapp{appendix: FJ transform}. 
Continuing the previous example of the axisymmetric configuration of a rotlet, when the rotlet is placed at $(x=0,y=0,z=r_0)$ with respect to the drop center, the coefficients for the expansion are
\begin{equation}
\begin{split}
&c_{j01}^{\displaced,-} =-\frac{i r_0^{j-1}}{4\viscratio} \sqrt{\frac{j(j+1)}{(2j+1)\pi}}\,, c_{j02}^{\displaced,+} = - \frac{i j r_0^{-j-2}}{4\sqrt{(2j+1)\pi}\viscratio}\,,
c_{j00}^{\displaced,+} =c_{j01}^{\displaced,+}= r_0^{-2j-1}c_{j01}^{\displaced,-}\,,
%&c_{j02}^{\displaced,+} = - \frac{i j r_0^{-j-2}}{4\sqrt{(2j+1)\pi}\viscratio}
\end{split}
\end{equation}
with all other coefficients equal to 0.

\item %Having found a spherical harmonic expansion for the singularity about the drop center, we can represent the solution for 
The total flow is a  superposition of the flow due to the singularity and perturbation due to the presence of the interface 
%inside and outside the droplet arr
% as the displaced singularity together with growing homogeneous solutions of the Stokes equation. 
%Likewise the flow outside the drop can be written as the sum of decaying solutions to the Stokes equation:
\begin{equation}
\begin{split}
&\bv^{\ins} =   \bv^{\displaced}+\sumjmq (c_{jm\q}-c_{jm\q}^{\displaced,-})\bu_{jm\q}^+\,,\quad 
%\sum_{j=1}^\infty \sum_{m=-j}^j\sum_{s=0}^2  c_{jm\q}^{\displaced,+}(\bxz)\bu_{jm\q}^+(\bx)H(|\bx_0|-|\bx|)+c_{jm\q}^{\displaced,-}(\bxz)\bu_{jm\q}^-(\bx)H(|\bx|-|\bx_0|)+(c_{jm\q}-c_{jm\q}^\infty)\bu_{jm\q}^+\\
\bv^{\out} =    \sumjmq c_{jm\q}\bu_{jm\q}^-.
\end{split}
\end{equation}
The above form for the solution satisfies the continuity of velocity at the spherical droplet interface. 
The unknown coefficients $c_{jm\q}$ are determined using the stress balance at the interface.
The hydrodynamic tractions associated with the flow are expanded in vector spherical harmonics  \citep{vlahovska2015dynamics},
%%%%%%Fix - 38 vectyor harmonics
\begin{equation}
\begin{split}
&\bt \equiv (\bT^{\ins}-\bT^{\out}) \cdot \rhat =  \sum \tau_{jm\q}\by_{jm\q}\,,\\
%&\tau_{jm\q} = T_{jm\q}^-(c_{jm\q}) - \viscratio(T_{jm\q}^+(c_{jm\q}-c_{jm\q}^{\displaced,-})+T_{jm\q}^-(c_{jm\q}^{\displaced,-} ))\,,
%&\tau_{jm\q} = c_{jm\q}T_{jm\q}^- - \viscratio((c_{jm\q}-c_{jm\q}^{\displaced,-})T_{jm\q}^++c_{jm\q}^{\displaced,-} T_{jm\q}^-)\,,
&\tau_{jm\q} = \sum_{\q'=0}^2 c_{jm\q'}T_{\q\q'}^- - \viscratio((c_{jm\q'}-c_{jm\q'}^{\displaced,-})T_{\q\q'}^++c_{jm\q'}^{\displaced,-} T_{\q\q'}^-)\,,
\end{split}\label{eq:traction11}
\end{equation}
\end{enumerate}
%where the amplitudes  $\tau_{di\q}$ are listed in \refapp{appendix: spherical harmonics}.
 where the  $T_{\q\q'}^\pm$ are listed in \refapp{appendix: spherical harmonics}.
 
Next we discuss the solution in the case of a surfactant-free drop where the hydrodynamic tractions are balanced by the surface tension only and surfactant-covered drop where the surface traction also include the Marangoni stresses.

\subsubsection{Clean droplet}

For a surfactant-free droplet, the tangential tractions are continuous ($\q=0,1$) and the normal tractions($\q=2$) is balanced by surface tension,
%remaining boundary condition is the stress balance% of the surface tranctions% can be written in terms of spherical harmonics as
\begin{equation}
\tau_{jm\q}= (-2+j(j+1)) f_{jm} \delta_{\q,2}\label{eq: stressbalance}
\end{equation}
where $\delta_{j,k}$ is the Kronecker delta function.

%The end result is the coefficients $c_{jm\q}$ are expressed in terms of the coefficients for the displaced singularity $c_{jm\q}^\infty$ and the shape deformation $f_{jm}$ as follows.
Solving for $c_{jm\q}$ yields,
\begin{equation}
\begin{split}
c_{jm1} = \frac{\viscratio (2j+1)}{2j+1+(\viscratio-1)(j-1)}c_{jm1}^{\displaced,-}\,,\quad  c_{jmn} = \zeta_j^{-1}\left( -p_{jmn} f_{jm}+q_{jmn}\right)\ \quad (n=0,2),\\
\end{split}\label{eq: cjms}
\end{equation}
where
\begin{equation}
\begin{split}
p_{jm0} = &3\sqrt{j(j+1)}(j-1)(j+2)\left((2j+1)+(\viscratio-1)(j+1) \right)\\
q_{jm0} =& (2j+1)\viscratio (3 c_{jm2}^{\displaced,-} \sqrt{j (1 + j)} (\viscratio-1) \\
&+ c_{jm0}^{\displaced,-}(j(4j^2+6j-1)+\viscratio (j+1)(4j^2+2j-3))\\
p_{jm2} =&(j-1)j(j+1)(j+2)(2j+1)(\viscratio+1)\\
q_{jm2} =& (2j+1)\viscratio (3c_{jm0}^{\displaced,-} \sqrt{j(j+1)} (\viscratio-1) \\
&+ c_{jm2}^{\displaced,-} ((4j^3+6j^2-j+3)+\viscratio(4j^3+6j^2-j-6))\\
\zeta_j =&(2j-1)(2j+1)^2(2j+3)+(\viscratio-1)(2j+1)(8j^3+12j^2-2j-9)\\
&\hspace{70pt}+2(\viscratio-1)^2(j-1)(j+1)(2j^2+4j+3).
\end{split}\label{eq: cjms2}
\end{equation}

\subsubsection{Surfactant-covered droplet}

%For the surfactant-covered drop,
In this case, the surface incompressibility condition \refeq{Eq: surface incompressibility} gives a relation between the amplitudes of the tangential and radial velocity components
\begin{equation}
c_{jm2} = \frac{1}{2}\sqrt{j(j+1)}c_{jm0}
\end{equation}
The stress balance for the surfactant-covered drop is
\begin{equation}
\begin{split}
\tau_{jm\q} = (2 g_{jm} + (-2+j(j+1)) f_{jm})\delta_{s,2}- \sqrt{j(j+1)} g_{jm}\delta_{s,0}
\end{split}
\end{equation}
%where we have used the coefficients $c_{jm\q}^{\surfactant}$ instead of the $c_{jm\q}$ for the clean drop given above. 
The above produce the following coefficients for the velocity and tension
\begin{equation}
\begin{split}
&c_{jm0} = \frac{2 c_{jm2}}{\sqrt{j(j+1)}}\,,\quad c_{jm1}=  \frac{\viscratio (2j+1)}{2j+1+(\viscratio-1)(j-1)}c_{jm1}^{\displaced,-}\\
& c_{jm2}=\xi_j^{-1}\left( -p_{jm2} f_{jm}+q_{jm2}\right) \,,\quad  g_{jm} =\xi_j^{-1}\left(-p_{jmg} f_{jm}+q_{jmg} \right)\\
\end{split}\label{eq: cjms surf}
\end{equation}
where
\begin{equation}
\begin{split}
&p_{jm2} =(j-1)j(j+1)(j+2)\\
&q_{jm2}= (2j+1)\viscratio(\sqrt{j(j+1)}c_{jm0}^{\displaced,-}+(2j^2+2j-3)c_{jm2}^{\displaced,-})\\
&p_{jmg} =(j-1)(j+2)\left((2j+1)+(\viscratio-1)(j-1)\right)\\
&q_{jmg}  =-(2j+1)\viscratio \bigg[ \frac{(2j-1)(2j+1)(2j+3)+(\viscratio-1)(j-1)(4j^2+10j+9)}{\sqrt{j(j+1)}}c_{jm0}^{\displaced,-}\\
&\hspace{40pt} -\frac{2(2j-1)(2j+1)(2j+3)+(j-1)(8j^2+17j+12)}{j(j+1)}c_{jm2}^{\displaced,-}\bigg]\\
&\xi_j = (2j+1)(2j^2+2j-1)+(\viscratio-1)(j-1)(2j^2+5j+5).
\end{split}\label{eq: cjms2 surf}
\end{equation}

\subsection{Interface evolution and steady shape}
The evolution equation for the interface  \refeq{Eq: surface evolution} is \citep{vlahovska2015dynamics} 
\begin{equation}
\Ca \frac{d f_{jm}}{dt} = c_{jm2} = -  p_{jm2}f_{jm}+q_{jm2}
\end{equation}
For both clean and surfactant-covered droplets, the coefficient $p_{jm2}>0$ for all $j>1$ indicating that when the singularity is fixed in place with respect to the drop center the shape modes $f_{jm}, j>2$ relax to a steady state value  given by
\begin{equation}
f_{jm}^\steadyshape =\frac{q_{jm2}}{p_{jm2}}.\label{eq: eqshape1}
\end{equation}
For a surfactant-free drops at steady state,
% $f_{jm}^\steadyshape$, defined in Equation \refeq{eq: eqshape1}, 
the flow for modes $j>1$ 
%due to the interface ???
is then given by
\begin{equation}\begin{split}
&c_{jm0}^\steadyshape = \viscratio \frac{2j(j+1)c_{jm0}^\displaced-3\sqrt{j(j+1)}c_{jm2}^\displaced}{j(j+1)(\viscratio+1)}\,,\\
 &c_{jm1}^\steadyshape =  \frac{\viscratio (2j+1)}{2j+1+(\viscratio-1)(j-1)}c_{jm1}^{\displaced,-}\,,\\
&  c_{jm2}^\steadyshape = 0.
\end{split}\end{equation}
%{\col{$\infty$ or $\displaced,-$?? }}
%Is this the solution for a spherical drop, i.e., assuming normal velocity=0, and not enforcing the normal stress}}
The $j=1$ flows remain the same as in \refeq{eq: cjms2}. 
The flow corresponding to the steady shape interface, $f_{jm}^\steadyshape$ matches the solution for the flow of a non-deformable spherical drop, given by setting $c_{jm2}=0$ and $f_{jm} = 0$  in the tangential stress balance  \refeq{eq: stressbalance} and solving for $c_{jm0}$ and $c_{jm1}$.

For a surfactant-covered drop at steady state, $f_{jm}^\steadyshape$, the flow for $j>1$ reduces to
\begin{equation}\begin{split}
&c_{jm0}^\steadyshape = c_{jm2}^\steadyshape = 0\,,\\
&c_{jm1}^\steadyshape =  \frac{\viscratio (2j+1)}{2j+1+(\viscratio-1)(j-1)}c_{jm1}^{\displaced,-}\,,\\
&g_{jm}^\steadyshape = (2j+1)\viscratio \frac{-2c_{jm0}^{\undisplaced,-} \sqrt{j(j+1)}+3c_{jm2}^{\undisplaced,-}}{j(j+1)}\,,
\end{split}\end{equation}
matching the flow about a spherical  surfactant-covered spherical drop. Notably, if the active particle generates axisymmetric flow with  $c_{jm1}^{\displaced,-}=0$ the interface is immobilized.

\subsection{Drop migration}
The translational velocity of a drop is defined as the volume average velocity of the fluid inside the drop.
\begin{equation}
\bV = \frac{1}{V} \int_V  \bv^{\ins} dV \label{VolumeAverageVelocity1}
\end{equation}
where $V$ is the volume of the drop. Under the assumption that the flow is incompressible everywhere, $\bnab  \cdot \bv^{\ins} = 0$, the following relation holds: $\bnab \cdot(\bx\bv^{\ins}) = \bv^{\ins}$. Under the further assumption of the smoothness of $\bv^{\ins}$, we can relate the drop translation to the surface velocity 
%For a spherical drop, 
%{\color{red} Make sure all formulas correct - I wrote the general ones, the spherical drop comes in for the second equality of 3.26}
%For a drop,
\begin{equation}
%\bV= \frac{3}{4\pi} \int_S (\bv^\ins \cdot \rhat) \rhat dS \label{VolumeAverageVelocity2}
\bV= \frac{1}{V} \int_S (\bv^\ins \bx) \cdot\nhat dS \label{VolumeAverageVelocity2}
\end{equation}
where $S$ is the surface of the drop. For a spherical drop, $\nhat=\xhat =\rhat$ and $V=4\pi/3$.
%where $S$ is the surface of a unit sphere.

Some care must be taken when placing singularities inside a droplet. Both the incompressibility and smoothness conditions may no longer hold. 
For a Stokeslet, rotlet, and stresslet, the singularity may be removed in the sense that
\begin{equation}
\int_{V_\delta} \bv dV_{\delta} =O(\delta) \to 0 \text{  as  }\delta\to 0
\end{equation}
where $V_{\delta}$ is a ball of radius $\delta$ around the singularity and $\bv$ is the velocity of the Stokeslet, rotlet, or stresslet. 
This results in definition \refeq{VolumeAverageVelocity1} being well defined and relation \refeq{VolumeAverageVelocity2} to hold.

For higher order singularities, we have that the volume average velocity as defined in \refeq{VolumeAverageVelocity1} does not converge. Note however if we were to consider an active particle of finite size with   volume  $V_{active}$, we have a well defined volume average velocity of the fluid:
\begin{equation}
\bV = \frac{1}{V} \int_{V-V_{active}}  \bv^{\ins} d(V-V_{active}) \label{VolumeAverageVelocity3}
\end{equation}
This can be given in terms of surface velocity contributions from the drop surface, $S$, and active particle surface, $S_{AP}$,
\begin{equation}
%\bV= \frac{1}{V} \int_S (\bv^{\ins}\cdot\nhat) \rhat dS + \frac{1}{V} \int_{S_{AP}} (\bv^{\ins}\cdot \nhat_{AP}) \nhat _{AP}dS_{AP}\label{VolumeAverageVelocity4}
\bV= \frac{1}{V} \int_S (\bv^{\ins} \bx) \cdot\nhat dS + \frac{1}{V} \int_{S_{AP}} (\bv^{\ins} \bx)\cdot \nhat_{AP} \bx dS_{AP}\label{VolumeAverageVelocity4}
\end{equation}
where %$S_{AP}$ is the surface of the active particle and 
$\nhat_{AP}$ is the unit normal vector pointing into the active particle.

For the remainder of the paper we will only consider the drop velocity in terms of the contribution from the drop surface and define the velocity of the spherical drop as
\begin{equation}
\bV_{\drop} = \frac{3}{4\pi} \int_S (\bv_s\cdot \rhat) \rhat dS= \frac{3}{4\pi} \sum_{m=-1}^1 c_{1m2} \int_S \by_{1m2}dS.
%\bV_{\drop} = \frac{1}{V} \int_S (\bv_s \bx)\cdot \nhatdS= \frac{3}{4\pi} \sum_{m=-1}^1 c_{1m2} \int_S \by_{1m2}dS {\color{red}+O(\epsilon)}.
\end{equation}
The integral of the vector spherical harmonic function can be taken on the unit sphere to yield
\begin{equation}
\bV_{\drop} = \frac{3}{4\pi}\sqrt{\frac{2\pi}{3}}\bigg(c_{1,-1,2}\begin{bmatrix}1\\-i\\0\end{bmatrix}+\sqrt{2}c_{1,0,2}\begin{bmatrix}0\\0\\1\end{bmatrix}+c_{1,1,2}\begin{bmatrix}-1\\-i\\0\end{bmatrix}\bigg).\label{eq: Drop Velocity SphH}
\end{equation}
For a clean drop, substituting \refeq{eq: cjms} for $c_{1m2}$ into \refeq{eq: Drop Velocity SphH}, leads to
\begin{equation}
\bV_{\drop}=\frac{\viscratio}{2(2+3\viscratio)}\sqrt{\frac{3}{2\pi}}
\begin{bmatrix}\sqrt{2}(c_{1,-1,0}^{\displaced,-}-c_{1,1,0}^{\displaced,-})(\viscratio-1)+(c_{1,-1,2}^{\displaced,-}-c_{1,1,2}^{\displaced,-})(\viscratio+4) \\
-i\sqrt{2}(c_{1,-1,0}^{\displaced,-}+c_{1,1,0}^{\displaced,-})(\viscratio-1)-i(c_{1,-1,2}^{\displaced,-}+c_{1,1,2}^{\displaced,-})(\viscratio+4) \\
2c_{1,0,0}^{\displaced,-}(\viscratio-1)+\sqrt{2}c_{1,0,2}^{\displaced,-}(\viscratio+4)
\end{bmatrix}.\label{eq:dropvel}
\end{equation}
The clean drop can only move as a results of singularities that generate the '$1m2$' mode when displaced from the origin. A closer inspection of the transforms in \cite{felderhof1989displacement} reveals that only linear combinations of the Stokeslet, rotlet, stresslet, and source dipole can generate non-zero $c_{1m2}$ coefficient values and thus result in non-zero drop velocity.
Higher order singularities do not induce droplet translation.

For a surfactant-covered drop, substituting  \refeq{eq: cjms surf} into \refeq{eq: Drop Velocity SphH}, yield for the drop velocity 
\begin{equation}
\bV_{\drop} = \frac{\viscratio}{3\sqrt{6\pi}}\begin{bmatrix}
\sqrt{2}(c_{1,-1,0}^{\displaced,-}-c_{1,1,0}^{\displaced,-})+c_{1,-1,2}^{\displaced,-}-c_{1,1,2}^{\displaced,-}\\
-i(\sqrt{2}(c_{1,-1,0}^{\displaced,-}+c_{1,1,0}^{\displaced,-})+c_{1,-1,2}^{\displaced,-}+c_{1,1,2}^{\displaced,-}\\
\sqrt{2}c_{1,0,0}^{\displaced,-}+c_{1,0,2}^{\displaced,-}
\end{bmatrix}\label{eq:dropvelsurf}
\end{equation}
Although there are four types of singularities that can produce non-zero $c_{1m2}$ coefficients, we will see in the next section that only the Stokeslet can produce a non-zero migration for the surfactant-covered drop. 
%Furthermore, there is no dependence on the location $\b\Rdrop  = (\Rdrop ,\theta_0,\phi_0)$ of the singularity.

We note that  the computation of the drop velocities in \cite{sprenger2020towards} assumes that the  net force on the droplet is given by the surface integral of the outer tractions only.  This formulation for the force only holds if there are no singularities inside the drop, and leads to erroneous result for the  drop velocity.

\subsection{Trajectory of the active particle}
The flow generated in response to the internal singularity advects the active particle.
The resulting trajectory is

\begin{equation}
\begin{split}
&\frac{d \bx_0}{dt} = \tilde{V}_p \bm{\hat{p}}+\sumjmq (c_{jm\q}-c_{jm\q}^{\displaced,-})\bu_{jm\q}^+\\
&\frac{d\bm{\hat{p}}}{dt} = \frac{1}{2}\left( \sumjmq (c_{jm\q}-c_{jm\q}^{\displaced,-})\bnab  \times\bu_{jm\q}^+\right)\times \bm{\hat{p}}
\end{split}\label{Eq: Particle Evo Eq 2}
\end{equation}

\section{Results and discussion}
\label{sec:Results}

Here we discuss the flows generated and the drop migration induced by a Stokeslet, rotlet and axisymmetric force dipole. Notable results for higher order singularities are also given. We show that transient shape deformation has a significant impact on the particle trajectory.
%{\col{ Shape deformation is shown to have a significant impact on particle trajectory. TALK ???}} {\col{ undispalced vs dispaced superscript, 

\subsection{Stokeslet}
The flow due to a point force in the direction $\bR = (d_x,d_y,d_z)$ in the coordinate system centered at the singularity is given by coefficients
\begin{equation}
\begin{split}
c_{1,-1,0}^{\undisplaced} = \frac{d_x+i d_y}{4\sqrt{3\pi}\viscratio}, \,\, c_{1,0,0}^{\undisplaced} = \frac{d_z}{2\sqrt{6\pi}\viscratio}, \,\, c_{1,1,0}^{\undisplaced} = -\frac{d_x-i d_y}{4\sqrt{3\pi}\viscratio}\\
c_{1,-1,2}^{\undisplaced} = \frac{d_x+i d_y}{2\sqrt{6\pi}\viscratio}, \,\, c_{1,0,2}^{\undisplaced} = \frac{d_z}{2\sqrt{3\pi}\viscratio}, \,\, c_{1,1,2}^{\undisplaced} = -\frac{d_x-i d_y}{2\sqrt{6\pi}\viscratio}.
\end{split}
\end{equation}
Upon using the transforms in \cite{felderhof1989displacement}, the coefficients for the flow in the coordinate system centered at the drop are obtained (see \refapp{appendix: FJ transform} for details):
\begin{equation}
\begin{split}
&c_{jm0}^{\undisplaced,-} =\frac{r_0 ^{j-1}}{2(2j+1)\viscratio}\bigg[-\frac{(j-2)(d_{jm2}\sqrt{j(j+1)}+d_{jm0}(j+1))}{2j-1}\\
&\hspace{150pt}+\frac{j(d_{jm2}\sqrt{j(j+1)}+d_{jm0}(j+3))}{2j+3}r_0 ^2\bigg]\\
&c_{jm1}^{\undisplaced,-} =\frac{r_0 ^j}{(2j+1)\viscratio}d_{jm1}\\
&c_{jm2}^{\undisplaced,-} =\frac{r_0 ^{j-1}}{2(2j+1)\viscratio}\bigg[\frac{(j+1)(d_{jm2}j+d_{jm0}\sqrt{j(j+1)})}{2j-1}-\frac{(d_{jm2}j(j+1)+d_{jm0}\sqrt{j(j+1)}(j+3))}{2j+3}r_0 ^2\bigg]\\
\end{split}
\end{equation}
\begin{equation}
\begin{split}
&c_{jm0}^{\undisplaced,+} =\frac{r_0^{-j}}{2(2j+1)\viscratio}\bigg[\frac{(j+1)(d_{jm2}\sqrt{j(j+1)}-d_{jm0}(j-2))}{2j-1}-\frac{(j+3)(d_{jm2}\sqrt{j(j+1)}-d_{jm0}j)}{2j+3}r_0^{-2}\bigg]\\
&c_{jm1}^{\undisplaced,+} =\frac{r_0^{-j-1}}{(2j+1)\viscratio}d_{jm1}\\
&c_{jm2}^{\undisplaced,+} =\frac{r_0^{-j}}{2(2j+1)\viscratio}\bigg[\frac{d_{jm2}j(j+1)-d_{jm0}\sqrt{j(j+1)}(j-2)}{2j-1}+\frac{j(-d_{jm2}(j+1)+d_{jm0}\sqrt{j(j+1)}}{2j+3}r_0^{-2}\bigg]
\end{split} \label{eq: coef disp stokeslet}
\end{equation}
where the singularity is located at $\bxz = (r_0,\theta_0,\phi_0)$ in spherical coordinates and $d_{jm\q} =  \bR \cdot \by_{jm\q}^\ast(\theta_0,\phi_0)$. 
The expressions above have been simplified and are not valid for $\theta_0=0,\pi$ as a result.
The full expressions listed in \refapp{app: Stokeslet displaced} should  be used to evaluate the coefficients for $\theta_0=0,\pi$.

\subsubsection{Stokeslet in a surfactant-free droplet}
Substituting \refeq{eq: coef disp stokeslet} into \refeq{eq: cjms} and \refeq{eq: cjms2} gives the coefficients for the flow accounting for the confining interface
\begin{equation}
c_{jm1} = \zeta_j^{-1}\left(\frac{r_0^{j} d_{jm1}}{2j+1+(\viscratio-1)(j-1)}\right),\quad
c_{jmn} = \zeta_j^{-1}\left( -p_{jmn}f_{jm}+q_{jmn}\right)\quad (n=0,2)\\
\label{eq:homcoefstokeslet}
\end{equation}
where
\begin{equation}
\begin{split}
&q_{jm0} = \frac{r_0^{j-1}}{2}\bigg[d_{jm0}\bigg((2j+1)\left(-(j-2)(j+1)(2j+3)+j(j+3)(2j-1)r_0^2\right)\\
&\hspace{70pt}+2(\viscratio-1)(j+1)\left(-(j+1)(j^2-j-3)+(j-1)j(j+3)r_0^2\right)\bigg)\\
&\hspace{40pt}+d_{jm2} \sqrt{j(j+1)}\bigg((2j+1)\left(-(j-2)(2j+3)+j(2j-1)r_0^2\right)\\
&\hspace{70pt}+2(\viscratio-1)(j+1)\left(-j^2+j+3+(j-1)j r_0^2\right)\bigg)\bigg]\\
&q_{jm2} = -\frac{r_0^{j-1}}{2}\bigg[ d_{jm0}\sqrt{j(j+1)}\bigg((2j+1)\left(-(j+1)(2j+3)+(j+3)(2j-1)r_0^2\right)\\
&\hspace{70pt}+2(\viscratio-1)(j+1)\left(-j(j+2)+(j-1)(j+3)r_0^2\right) \bigg)\\
&\hspace{40pt}+d_{jm2}j(j+1)\bigg((2j+1)\left(-(2j+3)+(2j-1)r_0^2\right)\\
&\hspace{70pt}+2(\viscratio-1)\left(-j(j+2)+(j-1)(j+1)r_0^2\right) \bigg)\bigg]\\
\end{split}
\end{equation}
and $\zeta_j$, $p_{jm0}$ and $p_{jm2}$ are provided in \refeq{eq: cjms2}.

%Given a Stokeslet held in place at position $\bxz$ with respect to the frame of reference moving with the drop, the shape of the drop will relax to 
The steady state shape of a drop with a Stokeslet fixed at position $\bxz$ with respect to the center of the drop is
\begin{equation}
\begin{split}
f^\steadyshape_{jm} &= \frac{q_{jm2}}{p_{jm2}}, \,\, j\geq 2 .
\end{split}
\end{equation}
Asymptotically for large $j$, $f_{jm}^\steadyshape \sim r_0^{j-1} j^{-1}$. 
For a Stokelet near the center of the particle, the amplitude of the shape mode decay is a power law, $r_0^{j-1}$, while for a Stokeslet closer to the boundary, the amplitudes decay as $1/j$.

From \refeq{eq:homcoefstokeslet} and \refeq{eq:dropvel}, the steady migration velocity of a 
%steady shape 
drop enclosing a Stokeslet is
\begin{equation}
\bV_{\drop} = \frac{1}{4\pi(2+3\viscratio)}\bigg[(2\viscratio+3)\bR+r_0^2(-2\bR+(\bR\cdot\rhat)\rhat) \bigg].
\end{equation}
The direction of drop translation is in general misaligned with the direction of the Stokeslet;  the correction is proportional to the off-center location of the singularity, $r_0$.
%in general in the same direction as the Stokeslet, with a $r_0^2$ off-center correction that can introduce motion of the drop not aligned with that of the Stokeslet.
The results for the steady flow and  velocity of the drop agree with those derived in \citep{kree2021dynamics} for non-deformable droplets.

\begin{figure}
	\centering
	\begin{subfigure}{0.31\textwidth}
	\includegraphics[width=1\linewidth,trim={50pts 0 70pts 0},clip]{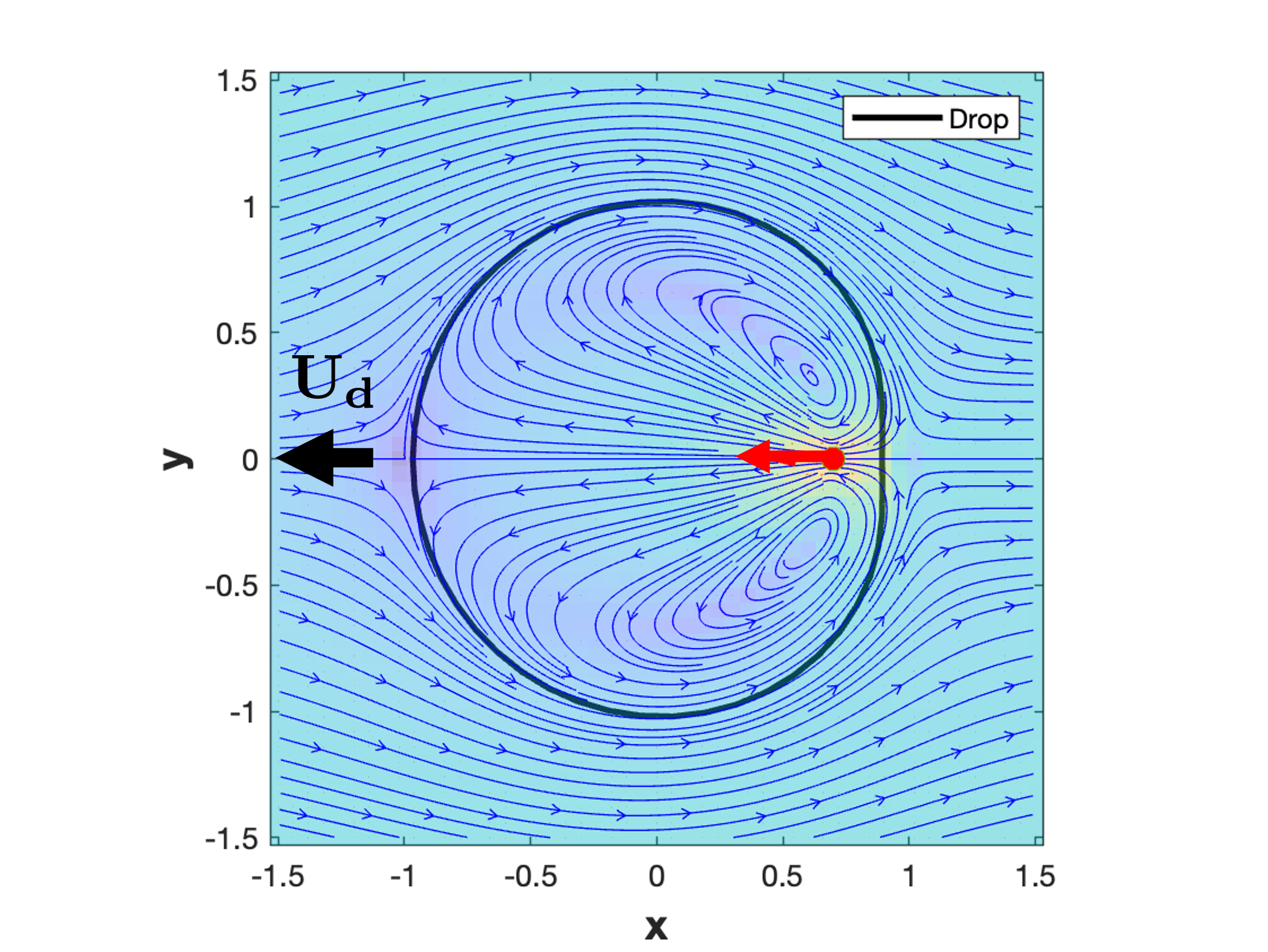}\caption{}
	\end{subfigure}
	\begin{subfigure}{0.31\textwidth}
	\includegraphics[width=1\linewidth,trim={50pts 0 70pts 0},clip]{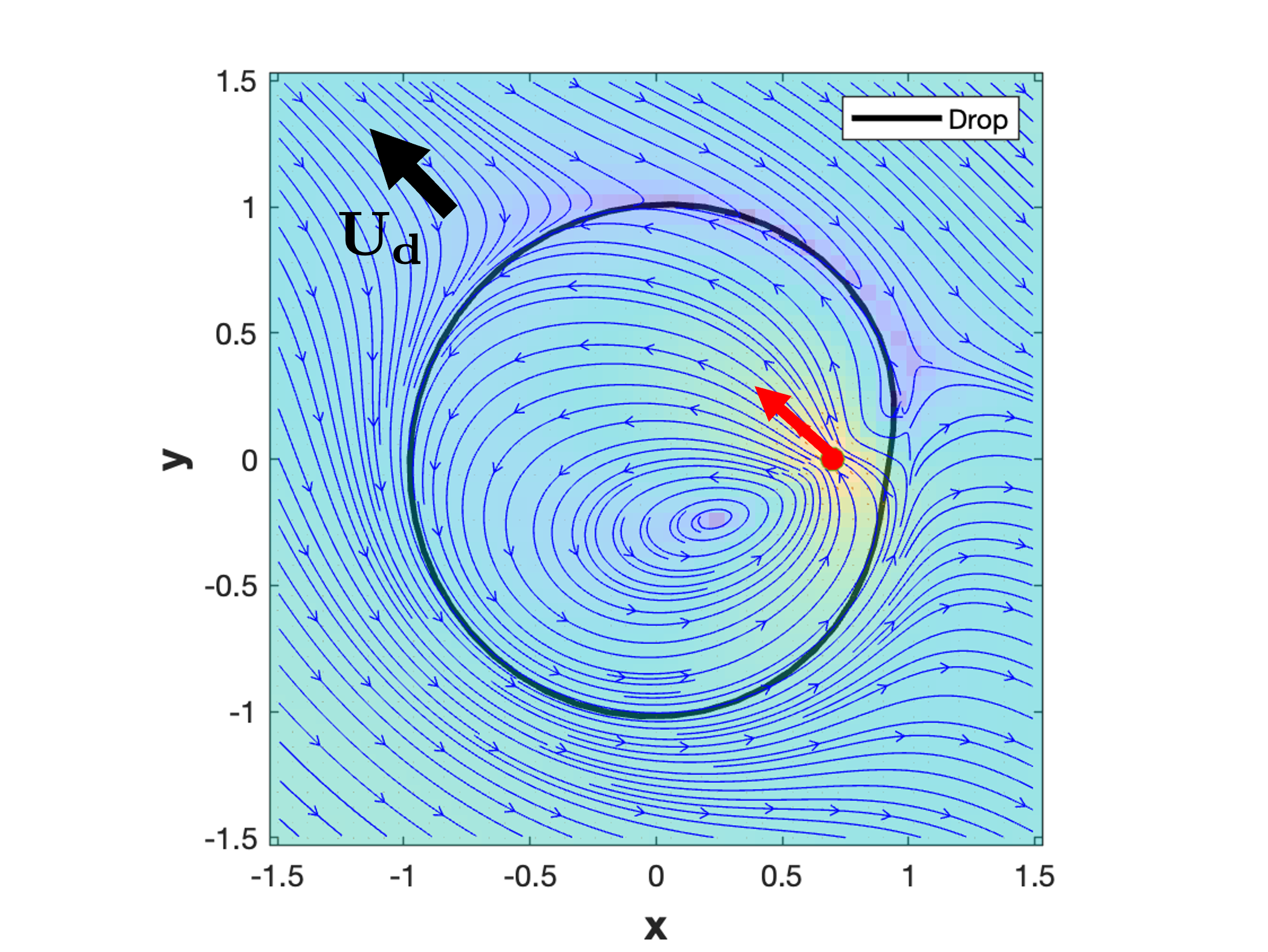}\caption{}
	\end{subfigure}
	\begin{subfigure}{0.31\textwidth}
	\includegraphics[width=1\linewidth,trim={50pts 0 70pts 0},clip]{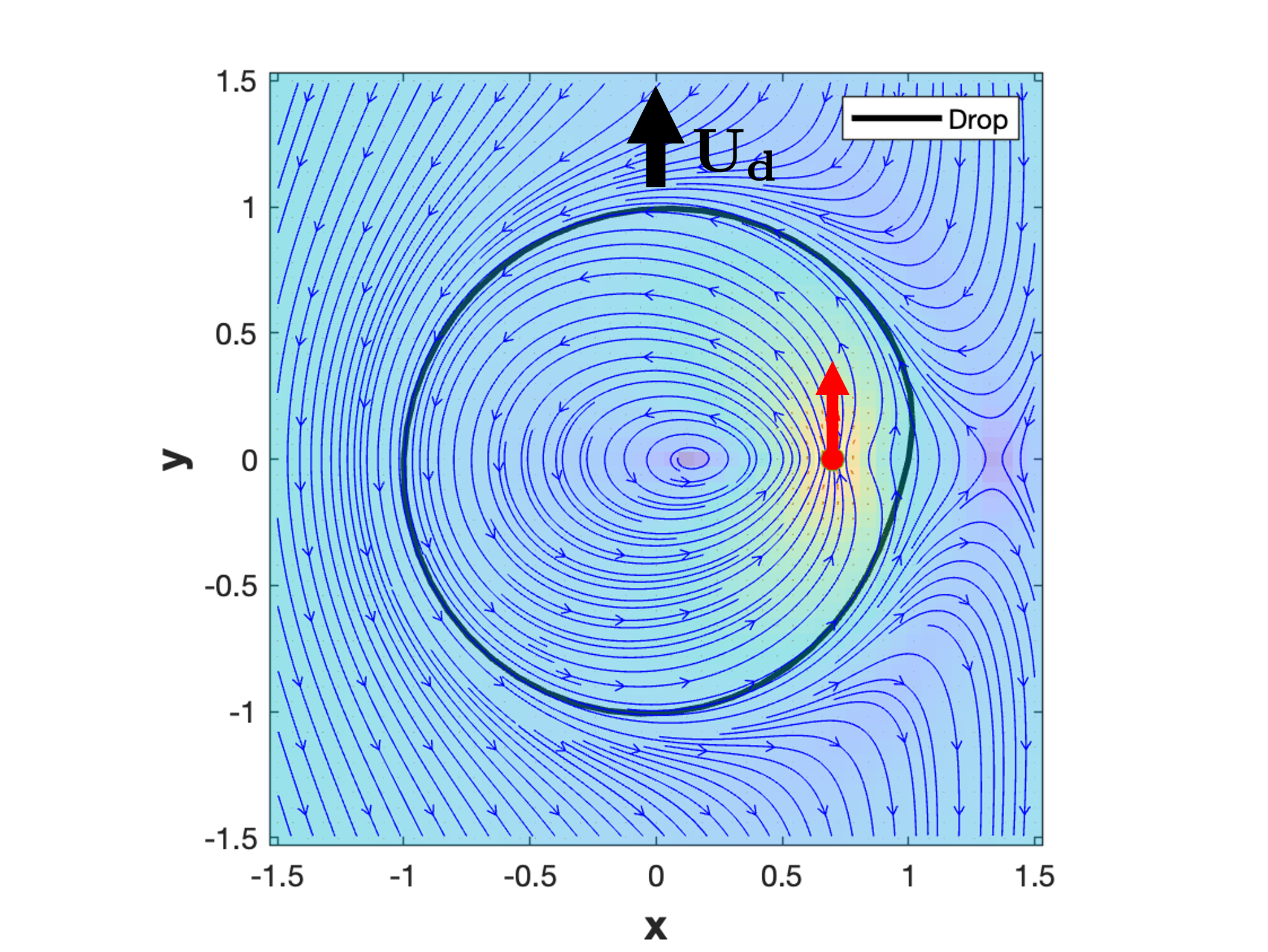}\caption{}
	\end{subfigure}
	\caption{\textbf{Stokeslet inside a surfactant-free droplet, $\bm{\bCa=.5, \viscratio=1}$.} Flow and drop deformation in response to a Stokeslet at position $(.7,0,0)$ with different orientations (a) $\bR = (-1,0,0)$, (b) $\bR = (-1/\sqrt(2),1/\sqrt(2),0)$, (c) $\bR = (0,1,0)$. Flows are given in the frame of reference moving with the droplet and the color scheme indicates the magnitude of the velocity. }%{\color{red} Take colorbar out and replace in caption - Color scheme indicates magnitude of flow. -Other option- include color bar,--- The colow indicates the $\ln|\bu|$ of the fluid velocity.}}
	\label{fig: Stokeslet Drop}
\end{figure}

\reffig{fig: Stokeslet Drop} shows the flow and steady shape of the interface for several configurations of the Stokeslet inside the surfactant-free droplet.  The interface of the drop is depressed inwards behind the Stokeslet and inflated outwards in front of it due to the Stokeslet pulling in fluid from behind and pushing it forward. The confinement leads to the circulation of the flow inside the drop and drop migration aligns mostly with the direction of the Stokeslet.

\subsubsection{Stokeslet in a surfactant-covered droplet}
For a surfactant-covered drop, the velocity  coefficients are obtained by substituting \refeq{eq: coef disp stokeslet} into \refeq{eq: cjms surf} and \refeq{eq: cjms2 surf}
\begin{equation}
\begin{split}
&c_{jm0} = \frac{2c_{jm2}}{\sqrt{j(j+1)}} \,,\quad  c_{jm1} = \frac{r_0 ^{j} d_{jm1}}{j+2+\viscratio(j-1)}\,,\\
& c_{jm2} =\xi_j^{-1}\left( -p_{jm2} f_{jm}+q_{jm2}\right)\,,\quad g_{jm} =  \xi_j^{-1}\left(-p_{jmg} f_{jm}+q_{jmg}\right)\,, \\
\end{split}
\end{equation}
where
\begin{equation}
\begin{split}
&q_{jm2} =-\frac{r_0 ^{j-1}}{2}\left(d_{jm0}\sqrt{j(j+1)}(j-1)(-(j+1)+(j+3)r_0^2)+d_{jm2} j(j+1)(-(j+1)+(j-1)r_0^2)\right)\\%\frac{d_{jm0}\sqrt{j(j+1)}(j^2-2j-1+r_0 ^2(j^2+2j-3))+d_{jm2}j(j+1)(-j-1+r_0 ^2(j-1))}{2(2j^3+3j^2+4+\viscratio(2j^3+3j^2-5))}\\
&q_{jmg} = -\frac{r_0^{j-1}}{2}\bigg[\frac{ d_{jm0}}{\sqrt{j(j+1)}}\bigg((2j+1)\left(-j(j+1)(2j+3)+(j+2)(j+3)(2j-1)r_0^2\right)\\
&\hspace{70pt}+2(\viscratio-1)(j-1)(j+1)\left(-(j^2+3j+3)+(j+2)(j+3)r_0^2 \right)\bigg)\\
&\hspace{50pt}+d_{jm2}\bigg((2j+1)\left(-j(2j+3)+(j+2)(2j-1)r_0^2\right)\\
&\hspace{70pt}+2(\viscratio-1)(j-1)(-(j^2+3j+3)+(j+1)(j+2)r_0^2)\bigg)\bigg] \\
%&\zeta = \frac{\viscratio r_0^{j-1}}{2\viscratio\sqrt{j(j+1)}(4-5\viscratio+j^2(2j+3)(\viscratio+1))}
\end{split}
\end{equation}
and $p_{jm2}$,$p_{jmg}$ and $\xi_j$ are given in \refeq{eq: cjms2 surf}.

This leads to steady state with shape mode amplitudes
\begin{equation}
\begin{split}
&f_{jm}^{\steadyshape} =-\frac{r_0 ^{j-1}}{2(j-1)j(j+1)(j+2)}\bigg[d_{jm0}\sqrt{j(j+1)}(j-1)(-(j+1)+(j+3)r_0^2)\\
&\hspace{100pt}+d_{jm2} j(j+1)(-(j+1)+(j-1)r_0^2)\bigg], \,\, j\geq 2.
\end{split}
\end{equation}
Asymptotically for large $j$, the amplitudes of the shape modes  decays as $f_{jm}^\steadyshape \sim r_0 ^{j-1} j^{-1}$, obeying the same general behavior as the clean drop case.

Evaluating \refeq{eq:dropvelsurf} with \refeq{eq:homcoefstokeslet}, shows that the drop velocity is 
\begin{equation}
\bV_{\drop}^{{\mathrm{\surfactant}}} = \frac{1}{6\pi}  \bR\,.
\end{equation}
Independent of the location of the Stokeslet inside the drop, the migration of the drop will be the same as a solid particle experiencing a force in the direction $\bR$ at its center. 
The flows and steady shapes for the surfactant-covered droplet are qualitatively similar to the flows and steady shape for surfactant-free droplets shown in \reffig{fig: Stokeslet Drop}.%

\subsection{Rotlet}
The flow due to a particle spinning in unbounded Stokes flow is given by the rotlet. 
%The expression for the  rotlet is given by
%\begin{equation}
%\bv^{\undisplaced}(\bx) = \frac{\bR\times\bx}{8\pi\viscratio |\bx|^3}
%\end{equation}
%chracterized by the axis of rotation $\bR$.
%
The coefficients for the rotlet in a coordinate system centered about itself is given by
\begin{equation}
c_{1,-1,1}^{\undisplaced} = \frac{-id_x+d_y}{4\sqrt{3\pi}\viscratio}, \,\, c_{1,0,1}^{\undisplaced} = -\frac{i d_z}{2\sqrt{6\pi}\viscratio}, \,\, c_{1,1,1}^{\undisplaced} = \frac{i d_x+d_y}{4\sqrt{3\pi}\viscratio}.
\end{equation}
%The full expressions for the coefficients of the rotlet at position $\bxz$ with respect to the center of the drop are given in \refapp{app: displaced rotlet}. 
%An equivalent expression, similar to those for the Stokeslet, can be derived using recursion relations for vector spherical harmonics.
The coefficients for the rotlet in a coordinate system centered at the drop is 
\begin{equation}
\begin{split}
&c_{jm0}^{\displaced,-} = -\frac{i j r_0^j}{2\viscratio(2j+1)}d_{jm1},\hspace{20pt}c_{jm2}^{\displaced,-} =\frac{i\sqrt{j(j+1)}r_0^j}{2\viscratio(2j+1)}d_{jm1}\\
&c_{jm1}^{\displaced,-} = -\frac{i(j+1)r_0^{j-1}}{2\sqrt{j(j+1)}(2j+1)\viscratio}(\sqrt{j(j+1)}d_{jm0}+jd_{jm2})\\
&c_{jm0}^{\displaced,+} = \frac{i(j+1)r_0^{-j-1}}{2\viscratio(2j+1)}d_{jm1}, \hspace{20pt} c_{jm2}^{\displaced,+} = \frac{i\sqrt{j(j+1)}r_0^{-j-1}}{2\viscratio(2j+1)}d_{jm1}\\
&c_{jm1}^{\displaced,+}  =\frac{i\sqrt{j(j+1)}r_0^{-j-2}}{2(j+1)(2j+1)\viscratio}(\sqrt{j(j+1)}d_{jm0}-(j+1)d_{jm2}).
\end{split}\label{eq:rotletdisplaced}
\end{equation}
Similar to the case of the Stokeslet, the expressions above cannot be evaluated for $\theta_0 = 0,\pi$ and the full expressions listed in \refapp{app: displaced rotlet} should be used in those cases.

\subsubsection{Rotlet in a surfactant-free droplet}

For the surfactant-free drop, substituting  \refeq{eq:rotletdisplaced} for the rotlet into \refeq{eq: cjms} and \refeq{eq: cjms2} yield the velocity coefficients
\begin{equation}
\begin{split}
& c_{jm0} =\zeta_j^{-1}\left(-p_{jm0} f_{jm}-\frac{i j(2j+3)r_0 ^j}{2}\left((2j-1)(2j+1)+2(\viscratio-1)(j-1)(j+1)\right)d_{jm1}\right)\\
&c_{jm1} = -\frac{i(j+1)r_0^{j-1}}{2\sqrt{j(j+1)}(2j+1+(\viscratio-1)(j-1))}(\sqrt{j(j+1)}d_{jm0}+j d_{jm2})\\
&c_{jm2} =\zeta_j^{-1}\left( -p_{jm2}f_{jm}+\frac{i\sqrt{j(j+1)}(2j+3)r_0^j}{2}\left((2j-1)(2j+1)+2(\viscratio-1)(j-1)(j+1)\right)d_{jm1}\right),
\end{split}\label{eq:homrotlet}
\end{equation}
where $p_{jm0},p_{jm2},\zeta_j$  are given in \refeq{eq: cjms2}.

The steady shape for a droplet with a rotlet fixed in place with respect to the drop center is
\begin{equation}
 f_{jm}^{\steadyshape}  = \frac{i r_0^j\sqrt{j(j+1)}(2j+3)\left((2j-1)(2j+1)+2(\viscratio-1)(j-1)(j+1)\right)}{2(j-1)j(j+1)(j+2)(\viscratio+1)}d_{jm1}, \,\, j\geq 2.
\end{equation}
Asymptotically for large $j$ , $f_{jm}^\steadyshape \sim r_0^j$. When the roltet is close to the center of the droplet the amplitude of the shape modes decay as $r_0^j$. For rotlets close to the interface, we see that the amplitude of the shape modes do not decay and are constant.

Substituting \refeq{eq:homrotlet} into and \refeq{eq:dropvel}, the velocity of the drop due to the rotlet is 
\begin{equation}
\bV_{\drop} = -\frac{5r_0}{8\pi(2+3\viscratio)}\bR\times \rhat \label{eq:rotletvel}
\end{equation}
An enclosed rotlet can only drive net motion of the drop if the axis of rotation does not point directly towards or away from the center of the drop. 
The velocity of the drop also scales linearly with the distance of the rotlet from the center of the drop with the direction of motion directed perpendicular to the axis of rotation of the rotlet and the line containing the singularity location and drop center.
A rotlet located in the center of a drop does not induce any drop migration.
\refeq{eq:rotletvel} agree with the velocity for a rotlet in a drop that can be derived from force dipole results given in \cite{kree2021dynamics}.

\begin{figure}
	\centering
	\begin{subfigure}{0.31\textwidth}
	\includegraphics[width=1\linewidth,trim={40pts 0 50pts 0},clip]{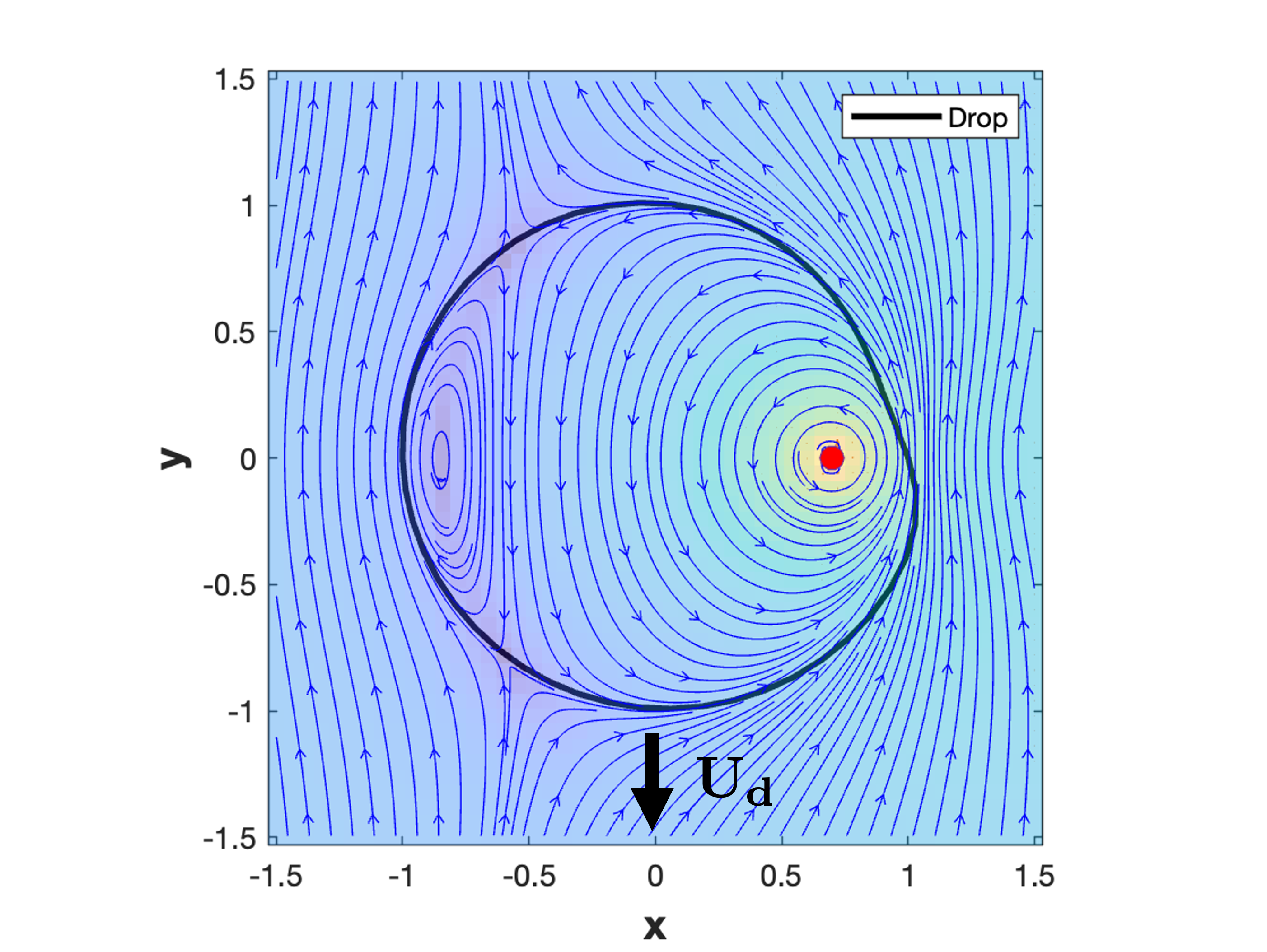}\caption{}\label{fig:Rotlet Dropa}
	\end{subfigure}
	\begin{subfigure}{0.31\textwidth}
	\includegraphics[width=1\linewidth,trim={40pts 0 50pts 0},clip]{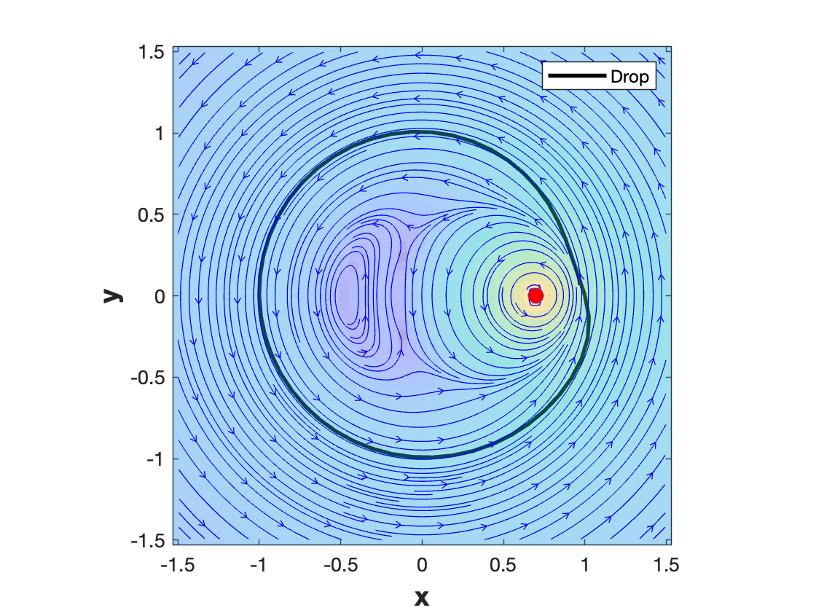}\caption{}\label{fig:Rotlet Dropb}
	\end{subfigure}
	\caption{\textbf{Rotlet inside  a droplet, $\bm{\bCa=.5,\viscratio=1}$.} The flow and shape deformation due to a rotlet placed at position $(.7,0,0)$ with orientation $\bR = (0,0,1)$ inside (a) a surfactant-free droplet and (b) a surfactant-covered droplet. Flows are in the frame of reference moving with the droplet and the color scheme indicates the magnitude of the velocity.}
	\label{fig:Rotlet Drop}
\end{figure}

An example of the flow induced by a rotlet inside a surfactant-free drop is shown in \reffig{fig:Rotlet Dropa}. 
The axis of rotation of the rotlet is $\xhat$ and it is spinning counterclockwise in the view of the x-y plane. % in \reffig{fig:Rotlet Dropa}. 
%The rotlet is spinning counterclockwise with respect to the view and the drop 
%in the lab frame 
%will move in the negative y-direction.
Both a depression and inflation of the interface is seen near the rotlet and the singularity induces drop migration in the negative y-direction.

\subsubsection{Rotlet in a surfactant-covered drop}
For the surfactant-covered drop, substituting \refeq{eq:rotletdisplaced} into \refeq{eq: cjms surf} and \refeq{eq: cjms2 surf} yields the following velocity coefficients 
\begin{equation}
\begin{split}
&c_{jm0} = \frac{2c_{jm2}}{\sqrt{j(j+1)}}\\
&c_{jm1} = -\frac{i(j+1)r_0^{j-1}}{2\sqrt{j(j+1)}(2j+1+(\viscratio-1)(j-1))}(\sqrt{j(j+1)}d_{jm0}+j d_{jm2})\\
&c_{jm2}= \xi^{-1}\left(-p_{jm2}f_{jm}+\frac{1}{2}i\sqrt{j(j+1)}(j-1)(2j+3)r_0^j d_{jm1}\right)\\
& g_{jm} = \xi^{-1}\left(-p_{jmg}f_{jm}+\frac{i(j+2)(2j+3)}{2\sqrt{j(j+1)}}\left((2j-1)(2j+1)+2(\viscratio-1)(j-1)(j+1) \right)r_0^jd_{jm1}\right).
\end{split}
\end{equation}
where $p_{jm2},p_{jmg},\xi$ are given in \refeq{eq: cjms2 surf}.

For a surfactant-covered drop containing a rotlet fixed in place with respect to the center of the drop, the steady shape is
\begin{equation}
 f_{jm}^{\steadyshape} = \frac{i(2j+3)r_0^j}{2(j+2)\sqrt{j(j+1)}}d_{jm1}.
\end{equation}
Asymptotically for large $j$, the amplitude of the shape modes is $f_{jm}^\steadyshape \sim r_0^j /j$. This differs by a factor of $j^{-1}$ from the case of the clean drop, with the difference highlighted as the rotlet is moved closer to the surface, i.e. $r_0  \to 1$.
Substituting \refeq{eq:homrotlet} into \refeq{eq:dropvelsurf}, reveals that the surfactant suppresses drop motion.

\reffig{fig:Rotlet Dropb} shows the flow induced by a rotlet in a steady shape surfactant-covered drop in the frame of reference moving with the drop.
% The flow for the surfactant covered drop differ from the flow for a clean drop qualitatively due to the lack of the translational $j=1$ modes. 
Compared to the flows produced by the rotlet enclosed in the surfactant-free droplet, the surfactant-covered droplet does not contain any translational flows associated with the $j=1$ modes.

\subsection{Axisymmetric stresslet}
The non-dimensionalized axisymmetric stresslet is given by
\begin{equation}
\bv^{\undisplaced}(\bx) = \frac{P}{8\pi\viscratio |\bx|^2}(-\bx+3(\bR\cdot\bx)^2\bx)
\end{equation}
where $P = \pm1$. $P=1$ corresponds to the flow produced by a pusher that expels fluid along $\bR$ and $P=-1$ corresponds to the flow of produced by pullers that expel fluid perpendicular to $\bR$ \citep{Lauga:2016,Saintillan:2018,Lauga_2020}.
The coefficients for an axisymmetric stresslet in a coordinate system centered at itself is
\begin{equation}
\begin{split}
&c_{2,-2,2}^{\undisplaced}  = \frac{P}{4\viscratio}\sqrt{\frac{3}{10\pi}}(d_x+i d_y)^2,
c_{2,-1,2}^{\undisplaced}  = \frac{P}{2\viscratio}\sqrt{\frac{3}{10\pi}}(d_x+i d_y)d_z,\\
&c_{2,0,2}^{\undisplaced}  = \frac{P}{4\viscratio\sqrt{5\pi}}(1-3 d_z^2),
c_{2,1,2}^{\undisplaced}  = \frac{P}{2\viscratio}\sqrt{\frac{3}{10\pi}}(d_x-i d_y)d_z, \\
&c_{2,2,2}^{\undisplaced}  = \frac{P}{4\viscratio}\sqrt{\frac{3}{10\pi}}(d_x-i d_y)^2.
\end{split}
\end{equation}
The expression for the coefficients of the axisymmetric stresslet at position $\bxz$ with respect to the drop center is given in \refapp{app: displaced stresslet}.

Using \refeq{eq:dropvel} for the clean drop and the coefficients for the flow due to confinement, obtained by substituing the results in \refapp{app: displaced stresslet} into \refeq{eq: cjms} and \refeq{eq: cjms2}, the droplet velocity induced by an axisymmetric stresslet located at $\bxz$ is
\begin{equation}
\bV_{\drop} = \frac{P r_0}{4\pi(2+3\viscratio)}(\rhat-3(\bR \cdot \rhat)\bR).
\end{equation}
The drop velocity scales linearily with the distance of the stresslet from the center of the drop. A stresslet at the center of a drop does not induce drop displacement. 
%The drop migration in the radial direction, $\rhat$, will change signs 

\begin{figure}
	\centering
	\begin{subfigure}{0.31\textwidth}
	\includegraphics[width=1\linewidth,trim={40pts 0 50pts 0},clip]{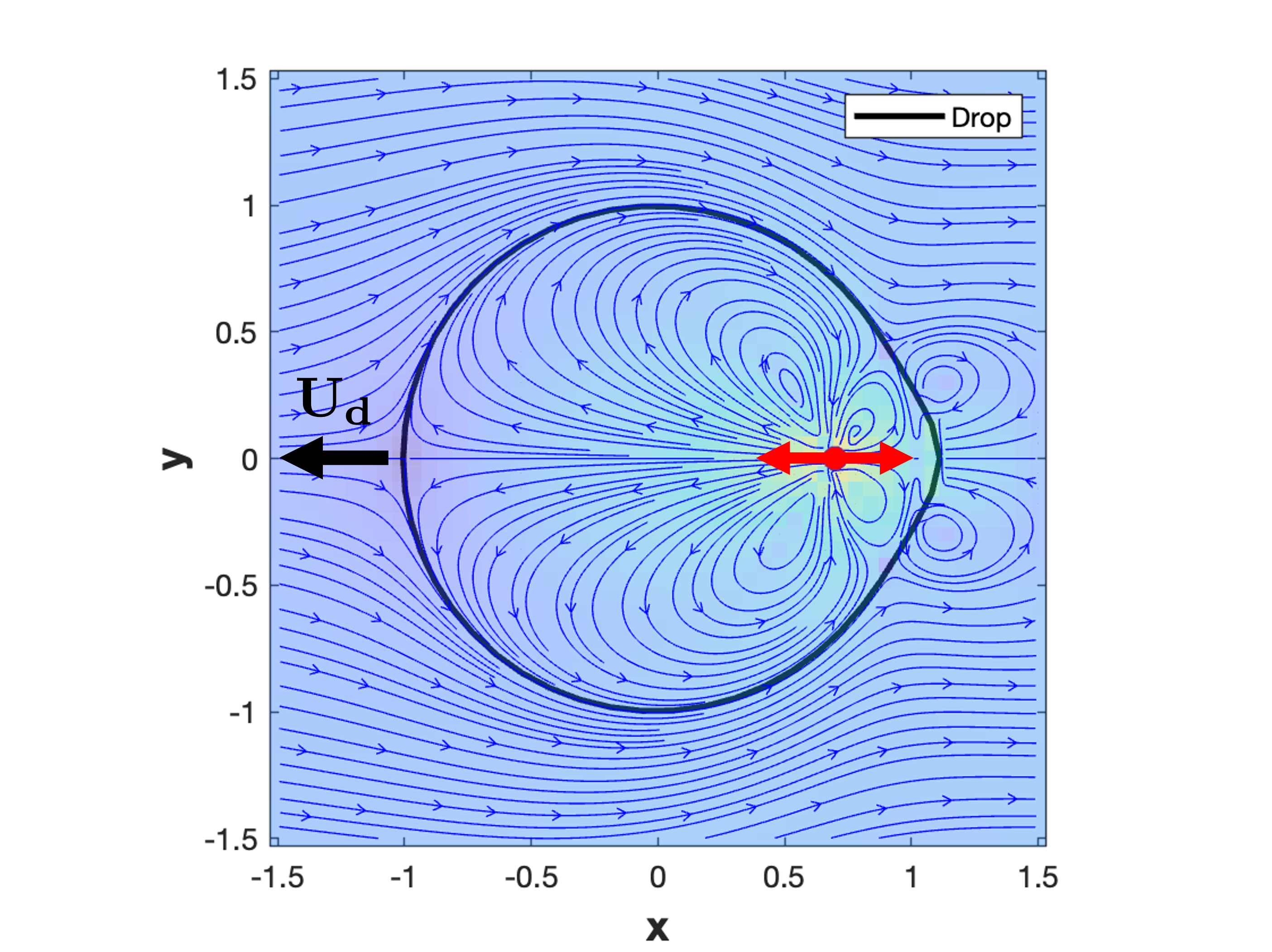}\caption{}
	\end{subfigure}
	\begin{subfigure}{0.31\textwidth}
	\includegraphics[width=1\linewidth,trim={40pts 0 50pts 0},clip]{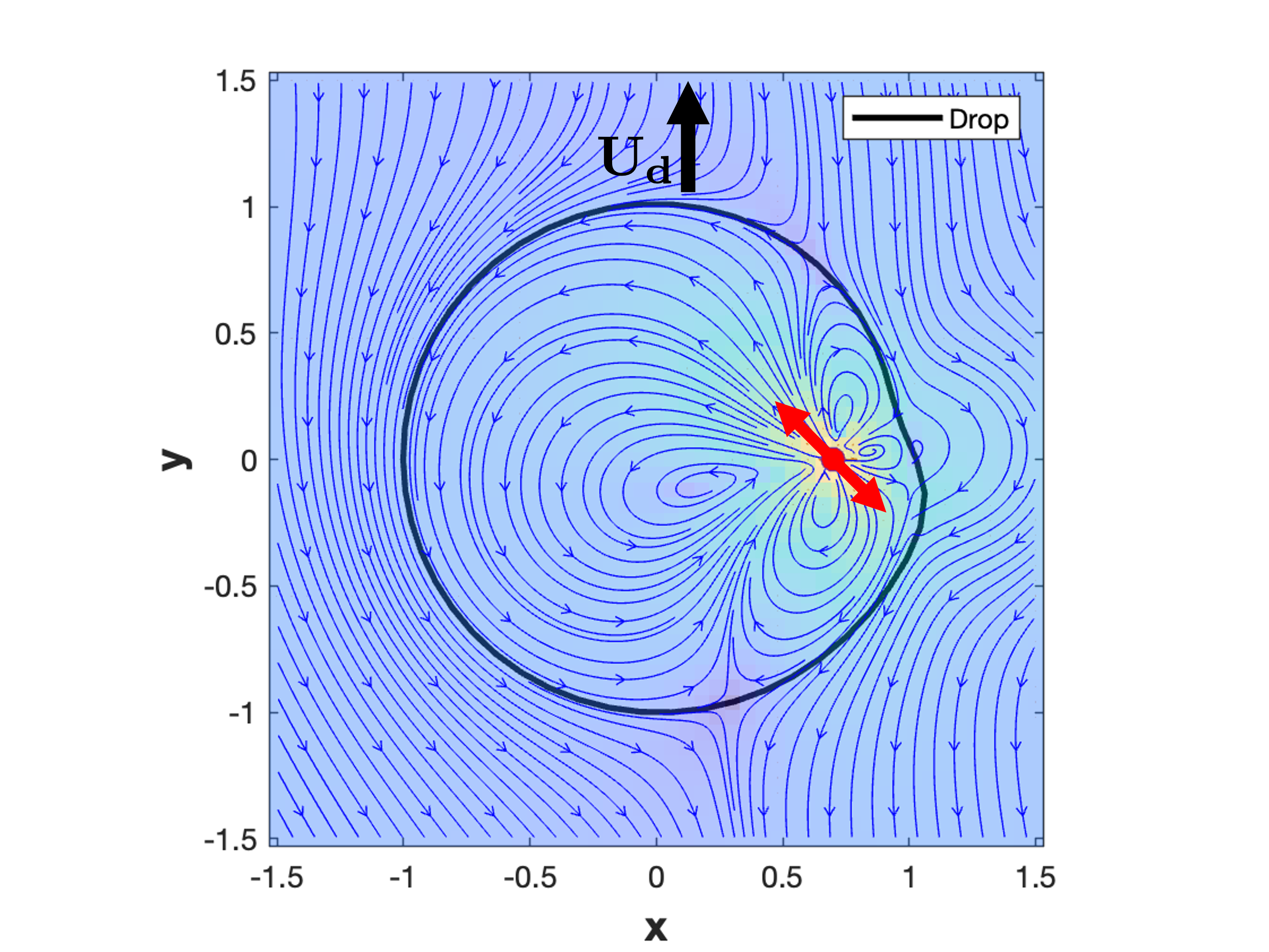}\caption{}
	\end{subfigure}
	\begin{subfigure}{0.31\textwidth}
	\includegraphics[width=1\linewidth,trim={40pts 0 50pts 0},clip]{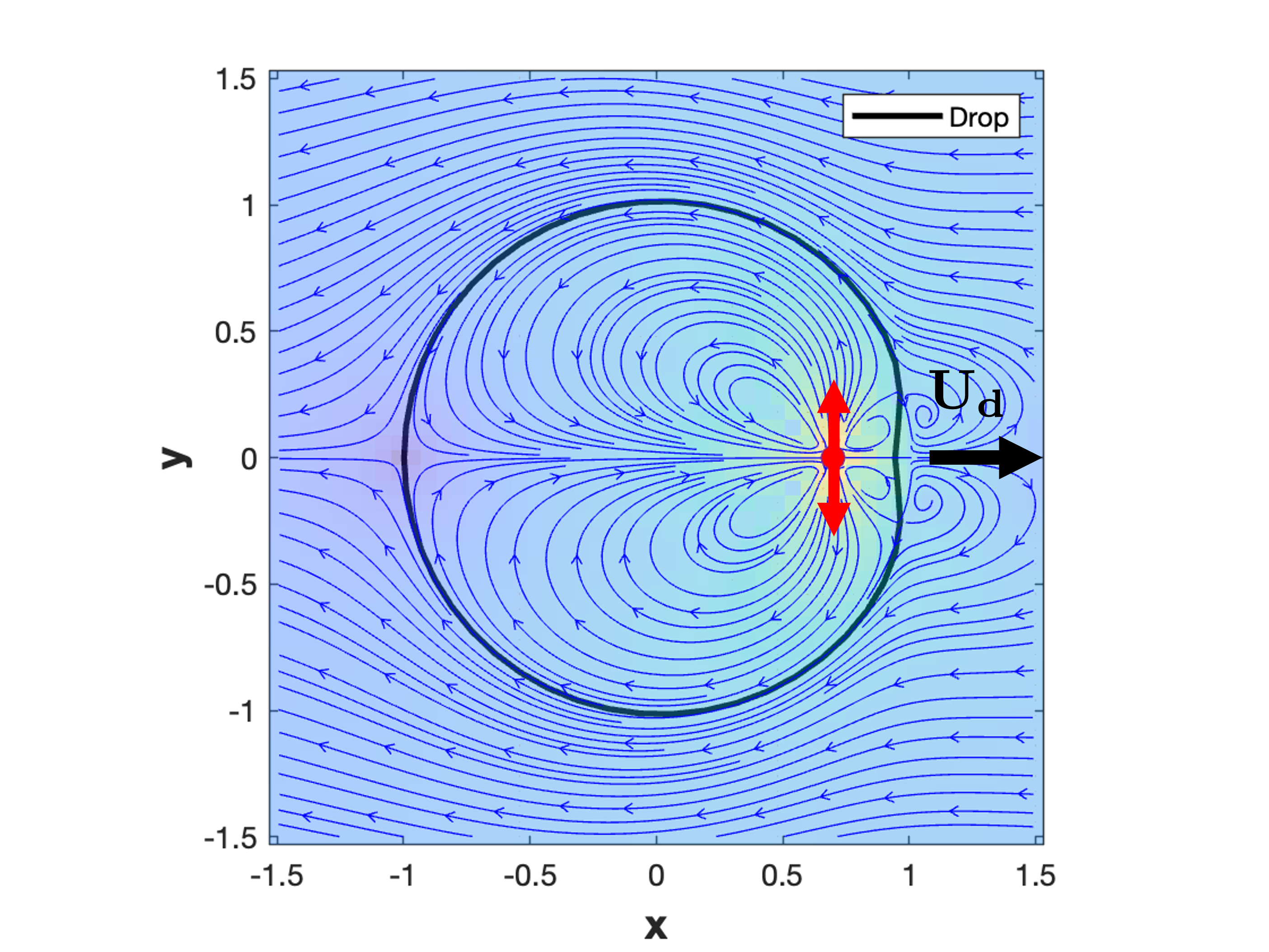}\caption{}
	\end{subfigure}
	\caption{\textbf{Axisymmetric stresslet inside a surfactant-free droplet, $\bm{\bCa=.2, \viscratio=1}$.} Flow and drop deformation in response to a axisymmetric stresslet located at $(.7,0,0)$ with different orientations (a) $\bR = (-1,0,0)$, (b) $\bR = (-1/\sqrt(2),1/\sqrt(2),0)$, (c) $\bR = (0,1,0)$. Flows are in the frame of reference moving with the droplet and the color scheme indicates magnitude of the velocity.}
	\label{fig: Stresslet Drop}
\end{figure}

\begin{figure}
	\centering
	\begin{subfigure}{0.31\textwidth}
	\includegraphics[width=1\linewidth,trim={40pts 0 50pts 0},clip]{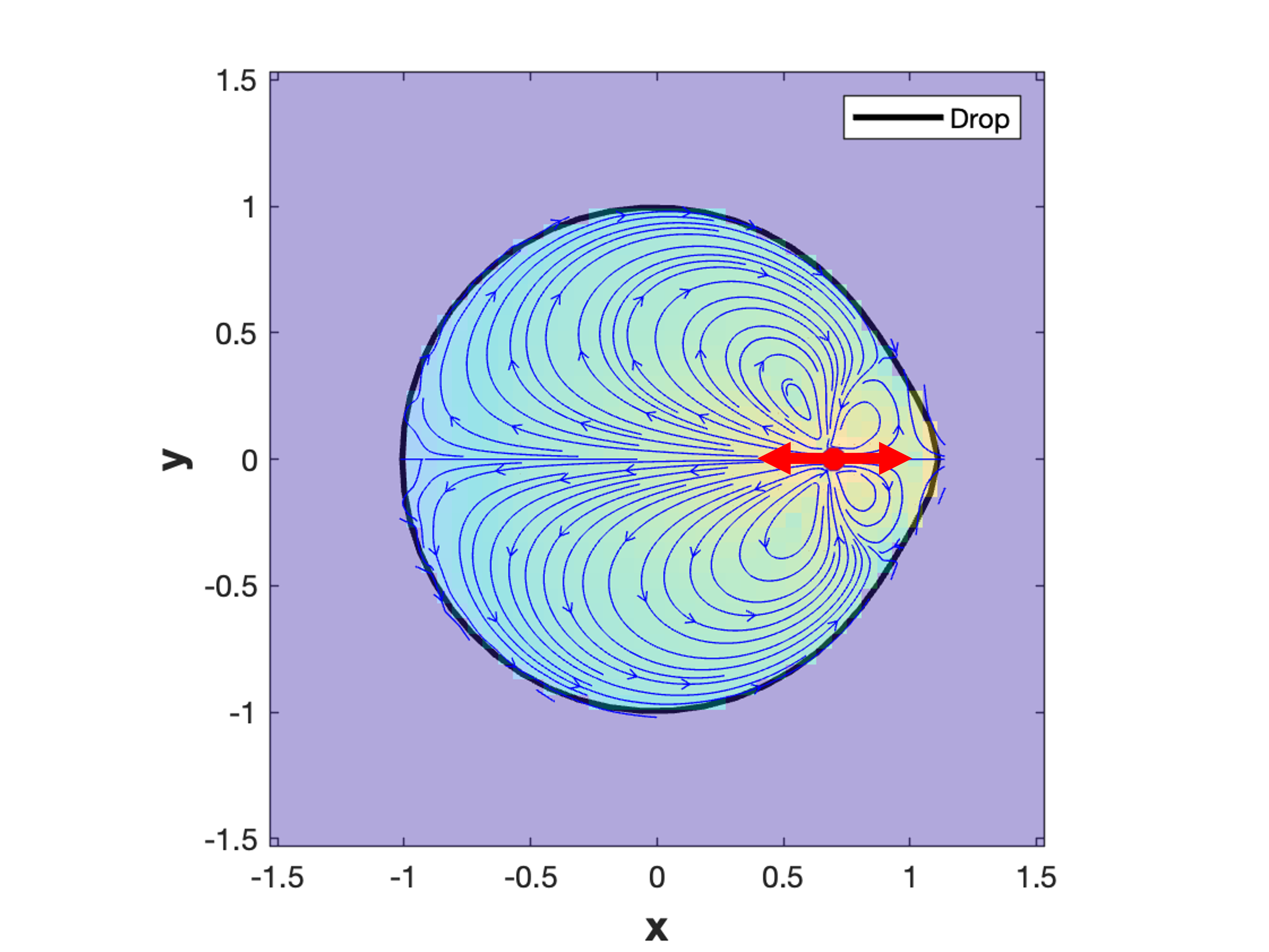}\caption{}\label{fig: stressletsurf1}
	\end{subfigure}
	\begin{subfigure}{0.31\textwidth}
	\includegraphics[width=1\linewidth,trim={40pts 0 50pts 0},clip]{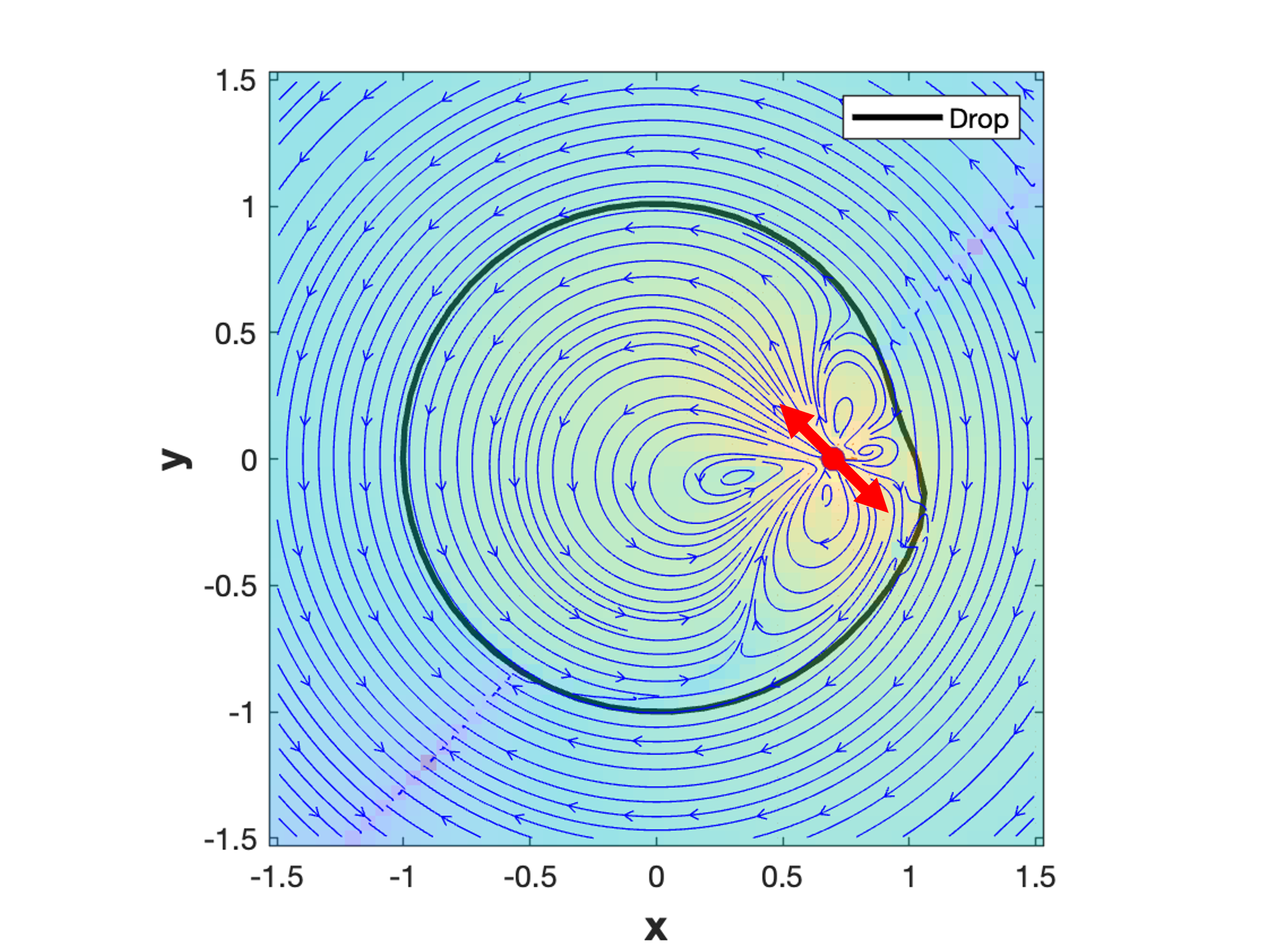}\caption{}
	\end{subfigure}
	\begin{subfigure}{0.31\textwidth}
	\includegraphics[width=1\linewidth,trim={40pts 0 50pts 0},clip]{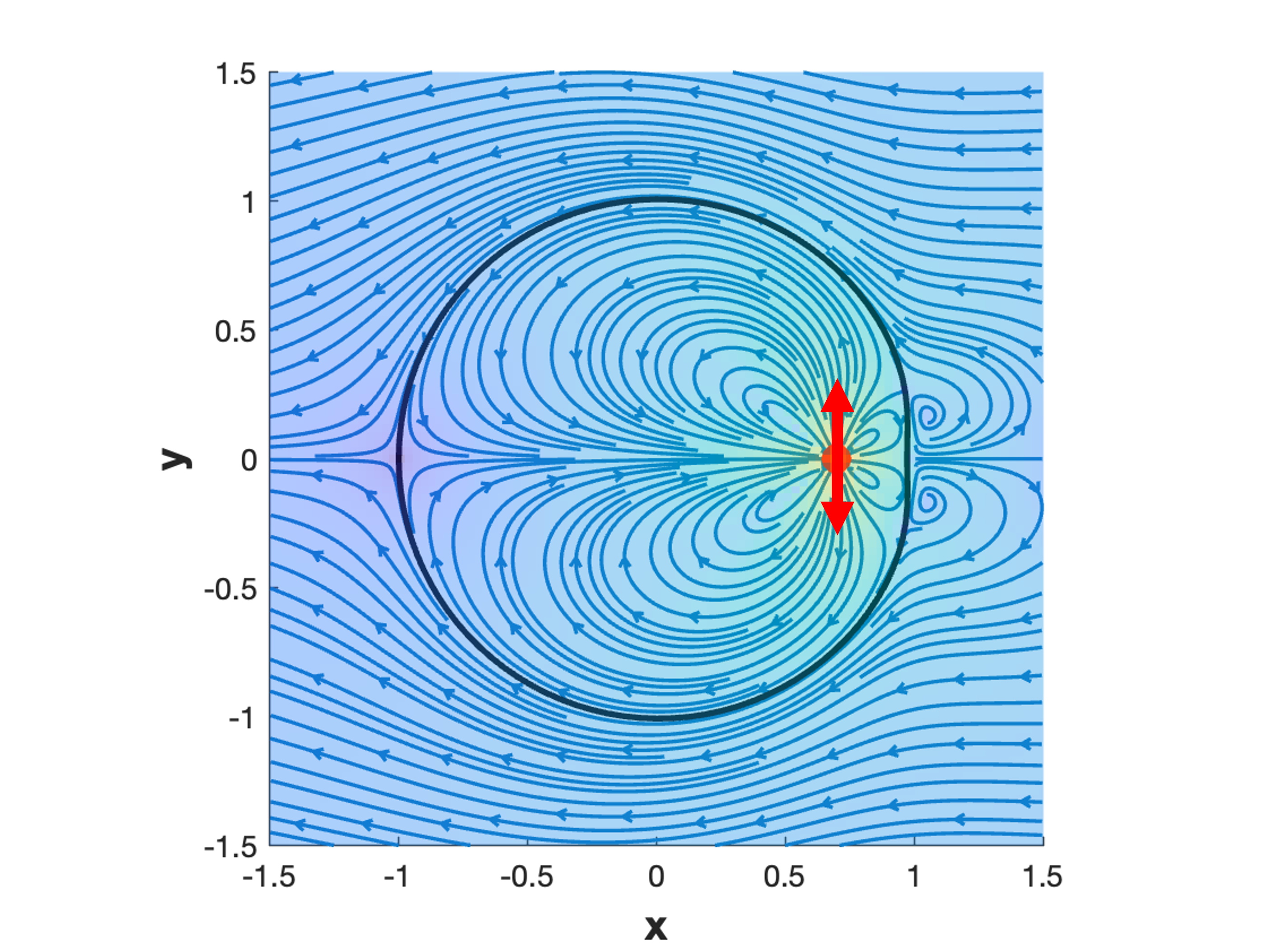}\caption{}
	\end{subfigure}
	\caption{\textbf{Axisymmetric stresslet inside a surfactant-covered droplet, $\bm{\bCa=.2, \viscratio=1}$.} Flow and drop deformation in response to a axisymmetric stresslet located at $(.7,0,0)$ with different orientations (a) $\bR = (-1,0,0)$, (b) $\bR = (-1/\sqrt(2),1/\sqrt(2),0)$, (c) $\bR = (0,1,0)$.  Flows are in the frame of reference moving with the droplet and the color scheme indicates magnitude of the flow.}
	\label{fig: Stresslet Surfactant Drop}
\end{figure}

For the surfactant-covered drop, substituting the results in \refapp{app: displaced stresslet} into \refeq{eq: cjms surf} and \refeq{eq: cjms2 surf}, to compute the coefficients of the flow due to confinement and using \refeq{eq:dropvelsurf} to compute the drop velocity, the drop migration is found to be suppressed for a surfactant-covered drop. 
\reffig{fig: Stresslet Drop} shows the steady shape and flow generated by an axisymmetric stresslet with $P=1$ in a surfactant-free droplet in the frame of reference of the drop. 
The interface of the drop is pushed out along the axis $\bR$ of the stresslet and is pulled in perpendicular to it.
\reffig{fig: Stresslet Surfactant Drop} shows the steady shape and flow generated by an axisymmetric stresslet with $P=1$ in a surfactant-covered drop. The axisymmetric configuration of the stresslet shown in \reffig{fig: stressletsurf1} does not induce motion of the fluid outside of the drop. There is no drop migration for all of the configurations of the stresslet inside of the surfactant-covered drop.

\subsection{Higher order singularities}
The flow and shape deformation for a droplet enclosing higher order singularities can be computed with the method described in the paper. 
%We do not provide the full solutions but they may be obtained for the following singularities using the coeffcients provided below.
The coefficients for the expansion of the source dipole and the rotlet dipole in a coordinate system centered at the singularity are provided below.
The full solution can be obtained from following the procedure in \refsec{sec: Solution Method}.

Source Dipole:
\begin{equation}
\begin{split}
&\bv^{\undisplaced}(\bx) = \frac{1}{8\pi\viscratio r^3}\bigg(-\bR+3\frac{\bR\cdot\bx}{r^2}\bx\bigg),\\
&c_{1,-1,0}^{\undisplaced}  = -\frac{d_x+id_y}{4\sqrt{3\pi}\viscratio}, c_{1,0,0}^{\undisplaced}  = -\frac{d_z}{2\sqrt{6\pi}\viscratio}, c_{1,1,0}^{\undisplaced}  = \frac{d_x-id_y}{4\sqrt{3\pi}\viscratio},c_{j,m,2}^{\undisplaced} = -\sqrt{2}c_{j,m,0}
\end{split}
\end{equation}

Axisymmetric rotlet Dipole:
\begin{equation}
\begin{split}
&\bv^{\undisplaced}(\bx) = \frac{3\chi}{8\pi\viscratio}\frac{(\bR\cdot \bx)(\bR\times \bx)}{r^5}\\
&c_{2,-2,1}^{\undisplaced}  = -\frac{3 i\chi (d_x+id_y)^2}{8\sqrt{5\pi}\viscratio},\,\, c_{2,-1,1}^{\undisplaced}  = -\frac{3i \chi(d_x+id_y)d_z}{4\sqrt{5\pi}\viscratio}\\
&c_{2,0,1}^{\undisplaced}  = \frac{\sqrt{3}i\chi (1-3 d_z^2)}{4\sqrt{10\pi}\viscratio},\,\, c_{2,1,1}^{\undisplaced}  = \frac{3i\chi(d_x-i d_i)d_z}{4\sqrt{5\pi}\viscratio}\, c_{2,2,1}^{\undisplaced }= -\frac{3i\chi (d_x-i d_y)^2}{8\sqrt{5\pi}\viscratio}
\end{split}
\end{equation}
where $\chi = \pm1$ indicates the rotation of the rotlets with respect to the swimming direction. For example, for swimming bacteria whose flagellar filaments rotate behind the cell in a counter-clockwise direction while the body counter rotates, $P=1$ \citep{Lauga_2020}.

In both the surfactant-free and surfactant-covered drops, the velocity of the drop is related to the amplitude of the $'1m2'$ modes in the expansion for the unbounded singularity in the coordinate system centered at the drop (see  \refeq{eq:dropvel} and \refeq{eq:dropvelsurf} ).
A careful examination of the displacement theorems from \cite{felderhof1989displacement} show that only the Stokeslet, rotlet, axisymmetric stresslet, source dipole and their linear combinations can excite the '1m2' mode and possibly result in non-trivial drop velocity. 
%The drop velocity due to the Stokeslet, rotlet, and axisymmetric stresslet have already been explored previously. 

The last remaining singularity that can induce non-trivial drop velocity is the source dipole and for a surfactant-free droplet, it is given by
\begin{equation}
\bV_{\drop} = \frac{5}{4(2+3\viscratio)\pi}\bR.
\end{equation}
The velocity of the drop is independent of the location of the source dipole and only dependent on its orientation.
The migration due to the source dipole is relevant when examining squirmers with the aspect ratio between the radius of the active particle and radius of the drop approaching 1. 
Under this condition, the source dipole may contribute significantly to the droplet motility compared to the leading order stresslet. 
 Additionally, unlike the stresslet, the source dipole can induce migration of the particle when located at the center of the  drop.
 
 However for the surfactant-covered droplet, despite the unbounded singularity exciting the $'1m2'$ modes, the source dipole will not induce drop motility.
For all other singularities, we get that $c_{1m2}^{\displaced,-} = 0$ and as a result regardless of surfactant-free or surfactant-covered, the drop will not move.

\subsection{Impact of the flow due to the confining drop on the active particle trajectory}
The trajectories of active particles near boundaries differ from their counterparts in unbounded fluid. 
The flow due to the presence of the interface can advect and rotate the active particle as it travels through the droplet.
\begin{align*}
&\frac{d\bx_0}{dt} = \tilde{V}_\particle\phat+\bu_\particle, \quad \frac{d\phat}{dt} = \bomega_\particle \times \phat\\
&\bu_\particle = \sum_{j,m,s} (c_{jm\q}^{\undisplaced,-}-c_{jm\q})\bu_{jm\q}^+(\bX_p), \quad \bomega_\particle =  \frac{1}{2}\bigg[\sum_{j,m,s} (c_{jm\q}^{\undisplaced,-}-c_{jm\q})\bnab \times \bu_{jm\q}^+(\bX_p) \bigg]
\end{align*}
$\bu_\particle$ and $\bomega_\particle$ are the translational and rotational velocities of the flow due to the presence of the soft confining interface.
For a squirmer, the orientation of the stresslet is the same as the propulsion direction, $\bR = \phat$. 
Note that this is not the case for all active particles, in some cases the orientation of the singularity may not align with the propulsion direction and higher order singularities may have multiple direction vectors parametrizing it.

For an active particle in a 
%surfactant-free 
drop in steady shape, the translational and rotational dynamics correspond to those of an active particle inside a non-deformable spherical drop. 
%The deformability of the interface leads to a richer vartiety of dynamics for the particle trajectory.
\begin{figure}
\centering
\includegraphics[width=\linewidth]{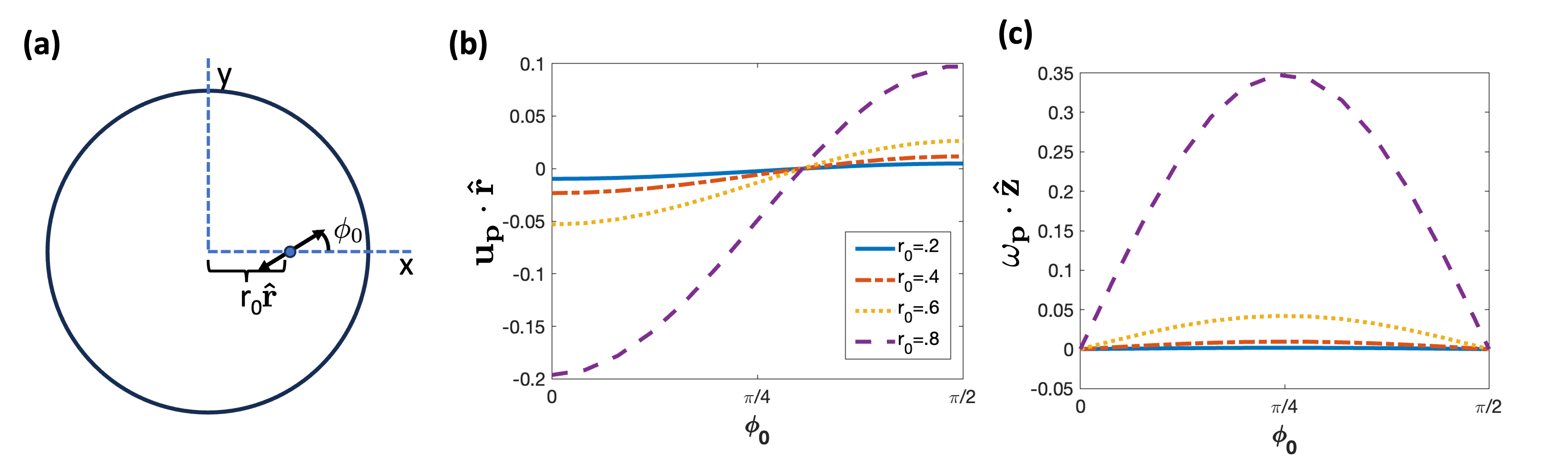}
\caption{\textbf{Feedback to active particle from the flow due to the presence of the steady shape interface.} $\bm{\bCa=.4}$, $\bm{\viscratio=3}$. (a).  Illustration of an axisymmetric stresslet at position $r_0 \rhat$ oriented on the xy-plane with angle $\phi_0$ with the radial direction $\rhat$. For this specific case we have $\rhat=\xhat$. (b) Translational velocity of the stresslet in the radial direction in response to the interface for steady shape droplets. The velocity is symmetric about $\phi_0 = \pi/2$. (c) Rotational velocity of the stresslet in response to the interface for steady shape droplets. The rotatinal velocity is anti-symmetric about $\phi=\pi/2$. }\label{fig: TranslationAFD}
\end{figure}
%Consider an active particle enclosed in a steady shape droplet. 
\reffig{fig: TranslationAFD} (a) illustrates a configuration of a stresslet, with $P=1$, inside  a droplet. The orientation relative to the radial direction, $\rhat$, is given by the angle $\phi_0$ and the distance from the center of the drop is given by $r_0$.
\reffig{fig: TranslationAFD} (b) and (c) shows the translational and angular velocity due to the flow generated by presence of the confining drop interface. The drop is surfactant-free with viscosity ratio $\viscratio=3$, capillary number $\Ca = .4$ and is in the steady state shape.
%This is equivalent to examining the feedback from the interface for the stresslet in a non-deformable spherical drop.
The correction to the particle swimming motion has a radial translational component that tends to repel  the particle away from the interface when its orientation is perpendicular to the interface;  when the particle orientation is parallel to the interface, the particle is attracted to the interface. 
The angular velocity of the active particle indicates turning in clockwise directions for $0<\phi_0<\frac{\pi}{2}$ and as a result the active particle will rotate to align parallel with the interface. 
The strength of both the translational and rotational velocity feedback increase as the stresslet approaches the interface.

\begin{figure}
\centering
\includegraphics[width=\linewidth]{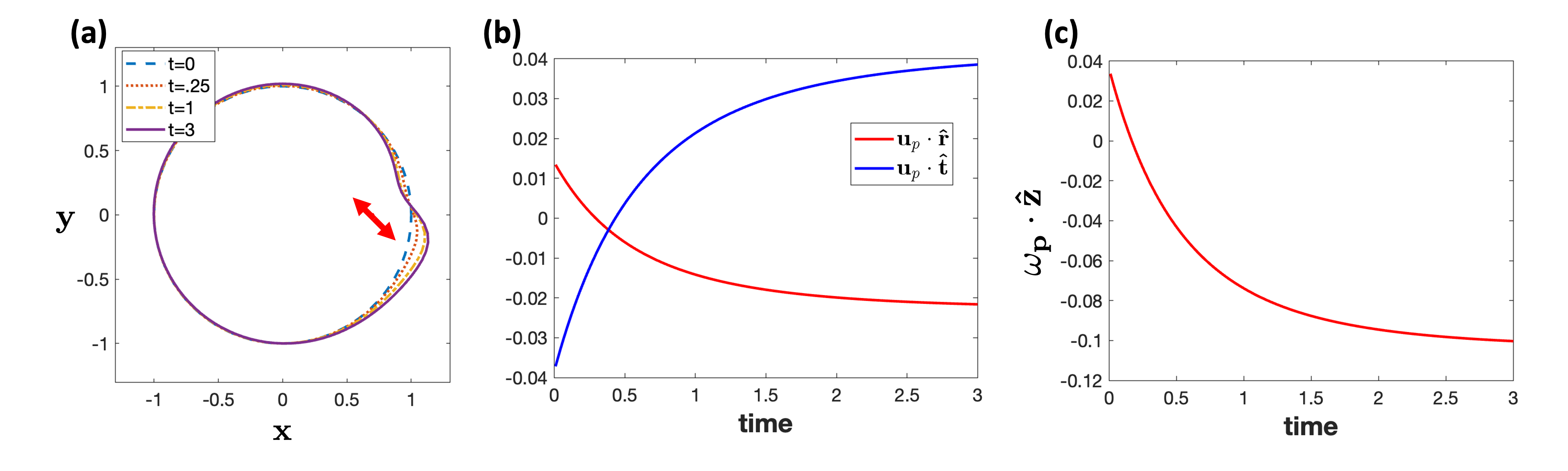}
\caption{\textbf{Translational and rotational velocity of a particle during  drop shape evolution towards steady state.} (a). A stresslet is placed at position $(.7,0,0)$ inside a clean spherical drop with viscosity ratio $\viscratio=3$ and capillary number $\Ca = .4$. The position and orientation of the singularity is held in place as the interface shape evolves towards the steady state shape. (b) The translational velocity of the stresslet due to the flow from the deforming interface in the radial and tangential direction. In this particlar case $\rhat = \xhat,\that =\yhat$. (c) The rotational rate of the stresslet due to the flow from the deforming interface.}\label{fig: TranslationAFD2}
\end{figure}

\reffig{fig: TranslationAFD2}  illustrates the effect of the transient drop shape on the translational and rotational velocity of a stresslet at location specified in \reffig{fig: TranslationAFD2} (a), 
%To illustrate the dependence of the response of the particle to the shape of the drop, consider the situation in \reffig{fig: TranslationAFD2} (a), where a stresslet with orientation
$\phi_0 =-\pi/4$ with the radial vector $\rhat$ and  $\Ca=.4,\mu=3$. 
%The position and orientation of the particle are held in place as the interface evolves towards the steady state shape.
The position and orientation of the stresslet are held static, the feedback  from the flow generated by the presence of the interface, $\bu_p$ and $\bomega_p$, are measured.
\reffig{fig: TranslationAFD2} (b) shows the evolution of the translational velocity, $\bu_p$,
%, due to the flow generated by the presence of the interface both f
components normal and tangential to the drop interface. 
We see that the  translation velocity $\bu_p$, reverses directions as the interface evolves towards the steady shape. Initially the particle is attracted, $\bu_p\cdot \rhat>0$, towards the interface but as the interface evolves, the particle is repelled.
 \reffig{fig: TranslationAFD2} (c) shows the evolution of the angular velocity feedback $\bomega_p$ from the interface. Similar to the translational velocity,  a reversal in the rotational direction is observed during the transient evolution of the drop shape. Initially the flow due to the presence of the soft interface works to align the stresslet  perpendicularly to the boundary but reverses to align the stresslet parallel to the interface as the drop shape  approaches steady state.
The coupling of the drop shape and trajectory of the active particle thus result in behavior not observable in non-deformable drops.

\section{Conclusions and outlook}
This paper describes a methodology to analytically solve for the flow generated by an active particle  enclosed by an initially spherical drop with either clean interface or interface covered with insoluble, nondiffusing surfactant.%and interfacial evolution for  a drop enclosing an active particle.
The approach models the active particle as an arbitrary Stokes flow singularity and assumes small interfacial deformation and surfactant redistribution.
%,  the case of the surfactant-covered drop,  Under further assumption of nearly uniform surfactant distribution, the analytical solution for surfactant-covered drops enclosing Stokes singularities were found.
The Stokeslet  is shown to induce drop translation in both surfactant-free and surfactant-covered drops while the rotlet, stresslet, and source dipole  induce non-zero drop migration only in a surfactant-free drop. In the surfactant-covered case, the Marangoni stresses immobilize the interface and arrest the external flow and drop motion, even though internal flow is present. The location and type of singularity influence strongly the direction of the drop motion, the drop shape, as well as 
%the drop shape are found to be important factors in
 the trajectory of the active particle inside. For example, the direction of drop translation is generally misaligned with the direction of the Stokeslet and drop shape is asymmetric. A pusher type stresslet tends to align parallel 
 %its extensional axis parallel
  to the interface and pull on it producing a ``dimple''. 

%Other directions
The analytical results provide useful insights into the physical mechanisms of droplet motility driven by encapsulated active particles. Future work will explore the dynamics of a swimming particle in a drop; it is expected that run-and-tumble locomotion will result in random shape fluctuations and non-trivial droplet trajectory. The methods developed in this paper can be applied to other complex interfaces such as lipid and polymer membranes (vesicles and capsules) \citep{vlahovska2015dynamics}. 
The approach can be extended to many particles
%The solution for multiple particles in a drop can be approximated via superposition of single particle solutions 
and can be used to study hydrodynamically  bound states in two particle systems such as those explored  in \citep{guo2024hydrodynamic} and collective behavior for many particle systems observed in the experiment e.g. \citep{park2022response} and \citep{kokot2022spontaneous}.

\section{Acknowledgements}

 This research was supported by NSF awards  DMR-2004926 and  DMS-2108502.
%%%%%%%%%%%%%%%%%%%%%%%%%%%%%%%%%%%%%%%%%%%%%%%%%%%%%%%%%%%%%%%%%%%%%%%%%%%%%%%%%%
%%%%%%%%%%%%%%%%%%%%%%%%%%%%%%%%%%%%%%%%%%%%%%%%%%%%%%%%%%%%%%%%%%%%%%%%%%%%%%%%%%

\appendix
\section{Spherical harmonics and fundamental solutions of the Stokes equation}
\label{appendix: spherical harmonics}
%\subsection{Definitions}
\label{Ap:vel fields}

The normalized scalar spherical harmonic are defined as 
\begin{equation}
\ysp_{jm}(\theta,\phi) = \bigg[\frac{2j+1}{4\pi} \frac{(j-m)!}{(j+m)!}\bigg] P_{j}^{m}(\cos(\theta))e^{i m \phi}\label{Def: sphericalharmonics}
\end{equation}
where $\theta$ and $\phi$ are the polar and azumithal angles in spherical coordinates, and $P_{j}^m$ are the associated Legendre polynomials.

The vector spherical harmonics are defined as
\begin{equation}
\begin{split}
&\by_{jm0} = \frac{1}{\sqrt{j(j+1)}}r\bnab _{\Omega} Y_{jm},\,\quad\by_{jm1} = -i \rhat \times \by_{jm0},\,\quad\by_{jm2} = Y_{jm} \rhat
\end{split}\label{Def: vsphericalharmonics}
\end{equation}
where $\bnab _{\Omega} $ denotes the angular part of the gradient operator. 

%\subsection{Fundamental set of solutions for the Stokes equation}
Following the definitions in \citep{vlahovska2015dynamics}, we list a  basis for solutions to the Stokes equations:
\begin{align}
&\bu^-_{jm0} = \frac{1}{2}r^{-j}(2-j+jr^{-2})\by_{jm0}+\frac{1}{2}r^{-j}\sqrt{j(j+1)}(1-r^{-2})\by_{jm2}\\
&\bu_{jm1}^- = r^{(-j-1)}\by_{jm1}\\
&\bu_{jm2}^- = \frac{1}{2}r^{-j}(2-j)\sqrt{\frac{j}{j+1}}(1-r^{-2})\by_{jm0}+\frac{1}{2}r^{-j}(j+(2-j)r^{-2})\by_{jm2}\\
&\bu_{jm0}^+ = \frac{1}{2}r^{j-1}(-(j+1)+(j+3)r^2)\by_{jm0}-\frac{1}{2}r^{j-1}\sqrt{j(j+1)}(1-r^2)\by_{jm2}\\
&\bu_{jm1}^+ = r^j\by_{jm1}\\
&\bu_{jm2}^+ = \frac{1}{2}r^{j-1}(j+3)\sqrt{\frac{j+1}{j}}(1-r^2)\by_{jm0}+\frac{1}{2}r^{j-1}(j+3-(j+1)r^2)\by_{jm2}.
\end{align}
On the unit sphere, the velocity fields reduce to
\begin{equation}
\bv^{\pm}_{jm\q} = \by_{jm\q}.
\end{equation}
Each of the basis fields are also incompressible
\begin{equation}
\bnab \cdot\bv^{\pm}_{jm\q} = 0.
\end{equation}
%\subsection{Hydrodynamic stresses}
Given a velocity of the following form
\begin{equation}
\bv = \sumjmq c_{jm\q}^\pm \bu_{jm\q}^{\pm},
\end{equation}
The $T_{jm\q}^\pm$ used in \refeq{eq:traction11} are defined via the radial traction of $\bu$:
\begin{equation}
\begin{split}
&\rhat \cdot (-p\bm{I}+\nabla \bu+(\nabla \bu)^T) = \sum_{j,m,s}\tau_{jm\q}^{\pm}\by_{jm\q}\\
&\tau_{jm\q}^\pm = \sum_{\q=0}^2 c_{jm\q'}^\pm T^\pm_{\q\q'}
\end{split}
\end{equation}
where

\begin{equation}
T_{\q\q'}^-=\begin{bmatrix}
-(2j+1)&0&3\sqrt{\frac{j}{j+1}}\\
0&-(j+2)&0\\
 3\sqrt{\frac{j}{j+1}}&0&-\frac{4+3j+2j^2}{j+1}
\end{bmatrix}
\end{equation}
and
\begin{equation}
T_{\q\q'}^+=\begin{bmatrix}
 (2j+1)&0&-3\sqrt{\frac{j+1}{j}}\\
0&(j-1)&0\\
  -3\sqrt{\frac{j+1}{j}}&0&\frac{3+j+2j^2}{j}
\end{bmatrix}
\end{equation}

The curl of the basis fields used in this paper are
\begin{align*}
&\bnab  \times\bu_{jm0}^+ = i r^{j}(2j+3)\by_{jm1}\\
&\bnab  \times\bu_{jm1}^+ = i r^{j-1} ((j+1)\by_{jm0}+\sqrt{j(j+1)}\by_{jm2})\\
&\bnab  \times\bu_{jm2}^+ = -ir^j\sqrt{\frac{j+1}{j}}(2j+3)\by_{jm1}
\end{align*}

\section{Felderhof-Jones transform}
\label{appendix: FJ transform}
The displacement theorems introduced in \cite{felderhof1989displacement} are used to relate the representation of a velocity field in terms of a fundamental solution basis centered at two different positions.
We note the last sentence before Section 4 in \cite{felderhof1989displacement} should also include $l=l'+1,\sigma'=0$ as an additional case for non-zero coefficients. This correction is important when calculating the expression for the stresslet about the center of a drop.

\subsection{Fundamental basis used in \citep{felderhof1989displacement} }
The fundamental basis for the solutions to the Stokes solution used in \citep{felderhof1989displacement}  differ from the one used throughout this paper. 
Define a normalization factor,
\begin{equation}
n_{jm} = \sqrt{\frac{4\pi}{2j+1} \frac{(j+m)!}{(j-m)!}}.
\end{equation}
Then define the following unnormalized vector spherical harmonics
\begin{equation}
\begin{split}
&\tilde{A}_{jm} = n_{jm}\left(j \by_{jm2} +\sqrt{j(j+1)}\by_{jm0}\right)\\
&\tilde{B}_{jm} = n_{jm}\left(-(j+1)\by_{jm2}+\sqrt{j(j+1)}\by_{jm0} \right)\\
&\tilde{C}_{jm} = n_{jm}  i \sqrt{j(j+1)}\by_{jm1}\\
\end{split}
\end{equation}
and a solution basis for the Stokes equation,
\begin{equation}\begin{split}
&\bu_{jm0}^{\FJ,+} = r^{j-1}\tilde{A}_{jm}\\
&\bu_{jm1}^{\FJ,+} = i r^j \tilde{C}_{jm}\\
&\bu_{jm2}^{\FJ,+} = r^{j+1}\left(\frac{(j+1)(2j+3)}{2j}\tilde{A}_{jm}+\tilde{B}_{jm}\right)\\
&\bu_{jm0}^{\FJ,-} = \frac{1}{(2j+1)^2}r^{-j}\left(\frac{j+1}{j(2j-1)}\tilde{A}_{jm}-\frac{1}{2}\tilde{B}_{jm}\right)\\
&\bu_{jm1}^{\FJ,-} = \frac{i}{j(j+1)(2j+1)}r^{-j-1}\tilde{C}_{jm}\\
&\bu_{jm2}^{\FJ,-} = \frac{j}{(j+1)(2j+1)^2(2j+3)}r^{-j-2}\tilde{B}_{jm}
\end{split}\end{equation}
we omit the pressure associated with the Stokes basis as it will not be necessary to conduct the displacement transform for the coefficients.

\subsection{Coefficient transformation between fundamental basis}
Given the flow
\begin{equation}
\bv = \sum_{j,m,s} c_{jm\q}^{\pm} \bu_{jm\q}^{\pm}= \sum_{j,m,s} a_{jm\q}^\pm\bu_{jm\q}^{\pm,FJ}
\end{equation}
the transformation between the coefficients for the representation of the flow in the two basis are 
\begin{equation}
\begin{split}
&c_{jm0}^- = \frac{n_{jm}}{2(2j+1)}\bigg(\sqrt{\frac{j+1}{j}}\frac{2-j}{2j-1}a_{jm0}^-+\sqrt{\frac{j}{j+1}}\frac{2j}{(2j+1)(2j+3)}a_{jm2}^- \bigg) \\
&c_{jm1}^- = \frac{n_{jm}}{(2j+1)\sqrt{j(j+1)}}a_{jm1}^-\\
&c_{jm2}^- = \frac{n_{jm}}{2j+1}\bigg(\frac{j+1}{2(2j-1)}a_{jm0}^--\frac{j}{(2j+1)(2j+3)}a_{jm2}^- \bigg)\\
&c_{jm0}^+ = n_{jm}\bigg(\sqrt{j(j+1)}a_{jm0}^++\frac{1}{2}(j+3)(2j+1)\sqrt{\frac{j+1}{j}}a_{jm2}^+\bigg)\\
&c_{jm1}^+ = n_{jm}\sqrt{j(j+1)}a_{jm1}^+\\
&c_{jm2}^+ = n_{jm}\bigg(j a_{jm0}^++\frac{1}{2}(j+1)(2j+1)a_{jm2}^+ \bigg)\label{eq: coordinatetransform}
\end{split}
\end{equation}

\subsection{Stokeslet}\label{app: Stokeslet displaced}
The following are the coefficients for the expansion of the Stokeslet located at $\bxz = (r_0,\theta_0,\phi_0)$ with orientation $(d_x,d_y,d_z)$ in a spherical coordinate system about the center of the drop. The coefficients are in the $\bu_{jm\q}^{FJ\pm}$ basis and can be converted to the basis used in this paper using \refeq{eq: coordinatetransform}. As a shorthand we abbreviate $\Yspp_{j,m} = y_{j,m}(\theta_0,\phi_0), \Sp_k = \sqrt{j+m+k}, \Tp_k = \sqrt{j-m+k}$.

\begin{flalign*}
&n_{jm}a_{jm0}^{\displaced,+} = \frac{(-1)^m r_0 ^{-j}}{2\viscratio(2j+1)}\bigg[&&\\
&\frac{-2(j+1)}{j(2j-1)^{3/2}\sqrt{2j+1}}\bigg(\frac{d_x+id_y}{2}\Tp_{-1}\Tp_0\Yspp_{j-1,-m-1}-d_z\Tp_0\Sp_0\Yspp_{j-1,-m}-\frac{d_x-id_y}{2}\Sp_0\Sp_{-1}\Yspp_{j-1,-m+1}\bigg)&&\\
&+\frac{1}{\sqrt{(2j+3)(2j+1)}}\bigg(\frac{d_x+id_y}{2}\Sp_1\Sp_2\Yspp_{j+1,-m-1}+d_z\Sp_1\Tp_1\Yspp_{j+1,-m}-\frac{d_x-id_y}{2}\Tp_1\Tp_2\Yspp_{j+1,-m+1} \bigg)\bigg]&&
\end{flalign*}

\begin{flalign*}
&n_{jm}a_{jm1}^{\displaced,+} =\frac{(-1)^m r_0^{-j-1}}{(2j+1)j(j+1)\viscratio}\bigg[ -\frac{d_x+i d_y}{2}\Tp_0\Sp_1\Yspp_{j,-m-1}+d_z m \Yspp_{j,-m}-\frac{d_x -i d_y}{2}\Sp_0\Tp_1\Yspp_{j,-m+1}\bigg] &&
\end{flalign*}

\begin{flalign*}
&n_{jm}a_{jm2}^{\displaced,+} = \frac{(-1)^m r_0^{-j-2} j}{(j+1)[(2j+1)(2j+3)]^{3/2}\viscratio}\\
&\hspace{30pt} \bigg[-\frac{d_x+i d_y}{2}\Sp_1\Sp_2\Yspp_{j+1,-m-1}-d_z\Sp_1\Tp_1\Yspp_{j+1,-m}+\frac{d_x-i d_y}{2}\Tp_1\Tp_2\Yspp_{j+1,-m+1} \bigg]&&
\end{flalign*}

\begin{flalign*}
&n_{jm}a_{jm0}^{\displaced,-} = \frac{(-1)^m r_0 ^{j-1}}{\viscratio}\sqrt{\frac{2j+1}{2j-1}}\\
&\hspace{30pt}\bigg[-\frac{d_x+i d_y}{2} \Tp_0\Tp_{-1}\Yspp_{j-1,-m-1}+ d_z\Sp_0\Tp_0\Yspp_{j-1,-m}+\frac{d_x-i d_y}{2}\Sp_0\Sp_{-1}\Yspp_{j-1,-m+1} \bigg]&&
\end{flalign*}

\begin{flalign*}
&n_{jm}a_{jm1}^{\displaced,-} =\frac{(-1)^m r_0^j}{\viscratio} \bigg[-\frac{d_x+i d_y}{2} \Tp_0\Sp_1\Yspp_{j,-m-1}+m d_z \Yspp_{j,-m}-\frac{d_x-i d_y}{2}\Sp_0\Tp_1\Yspp_{j,-m+1} \bigg]&&
\end{flalign*}

\begin{flalign*}
&n_{jm}a_{jm2}^{\displaced,-} = \frac{(-1)^mr_0^{j+1}\sqrt{2j+1}}{2\viscratio}\bigg[&&\\
&\frac{(j+1)(2j+3)}{j} \bigg(- \frac{d_x+i d_y}{2} \Tp_0\Tp_{-1}\Yspp_{j-1,-m-1}+d_z\Sp_0\Tp_0\Yspp_{j-1,-m}+\frac{d_x-i d_y}{2}\Sp_0\Sp_1\Yspp_{j-1,-m+1}\bigg)&&\\
&-\frac{2}{\sqrt{2j+3}}\bigg( \frac{d_x+i d_y}{2}\Sp_1\Sp_2\Yspp_{j+1,-m-1}+d_z\Sp_1\Tp_1\Yspp_{j+1,-m}-\frac{d_x-i d_y}{2}Tp_1\Tp_2\Yspp_{j+1,-m+1}\bigg)\bigg]&&
\end{flalign*}

\subsection{Rotlet}
\label{app: displaced rotlet}
The following are the coefficients for the expansion of the rotlet located at $\bxz = (r_0,\theta_0,\phi_0)$with axis of rotation $(d_x,d_y,d_z)$ in a spherical coordinate system about the center of the drop. The coefficients are in the $\bu_{jm\q}^{FJ\pm}$ basis and can be converted to the basis used in this paper using \refeq{eq: coordinatetransform}. As a shorthand we abreviate $\Yspp_{j,m} = y_{j,m}(\theta_0,\phi_0), \Sp_k = \sqrt{j+m+k}, \Tp_k = \sqrt{j-m+k}$.

\begin{flalign*}
&n_{jm}a_{jm0}^{\displaced,+}  = \frac{i(-1)^m r_0^{-j-1}}{2j(2j+1)\viscratio}\bigg[ -\frac{d_x+i d_y}{2}\Tp_0\Sp_1\Yspp_{j,-m-1}+m \Yspp_{j,-m}-\frac{d_x-i d_y}{2}\Sp_0\Tp_1\Yspp_{j,-m+1}\bigg]&&\\
&n_{jm}a_{jm1}^{\displaced,+}  = \frac{i(-1)^m r_0^{-j-2}}{2(j+1)\viscratio \sqrt{(2j+1)(2j+3)}}\\
&\hspace{30pt}\bigg[-\frac{d_x+i d_y}{2}\Sp_1\Sp_2\Yspp_{j+1,-m-1}-d_z\Sp_1\Tp_1\Yspp_{j+1,-m}+\frac{d_x-i d_y}{2}\Tp_1\Tp_2\Yspp_{j+1,-m+1} \bigg]&&\\
&n_{jm}a_{jm2}^{\displaced,+}  = 0
\end{flalign*}

\begin{flalign*}
&n_{jm}a_{jm0}^{\displaced,-} = 0\\
&n_{jm}a_{jm1}^{\displaced,-}  = \frac{i(-1)^m r_0^{j-1}}{2\viscratio}(j+1) \sqrt{\frac{2j+1}{2j-1}}\\
&\hspace{30pt}\bigg[\frac{d_x+i d_y}{2}\Tp_0\Tp_{-1}\Yspp_{j-1,-m-1}-d_z\Tp_0\Sp_0\Yspp_{j-1,-m}-\frac{d_x-i d_y}{2}\Sp_0\Sp_{-1}\Yspp_{j-1,-m+1} \bigg]&&\\
&n_{jm}a_{jm2}^{\displaced,-}  = \frac{i (-1)^m (2j+1)(2j+3) r_0^j}{2j\viscratio} \sqrt{\frac{2j+1}{2j-1}}\\
&\hspace{30pt}\bigg[\frac{d_x+i d_y}{2}\Tp_0\Tp_1\Yspp_{j,-m-1}-m \Yspp_{j,-m} +\frac{d_x-i d_y}{2}\Sp_0\Sp_1\Yspp_{j,-m+1}\bigg]&&
\end{flalign*}

\subsection{Axisymmetric stresslet}
\label{app: displaced stresslet}
The following are the coefficients for the expansion of the axisymmetric stresslet located at $\bxz = (r_0,\theta_0,\phi_0)$with orientation $(d_x,d_y,d_z)$ in a spherical coordinate system about the center of the drop.. The coefficients are in the $\bu_{jm\q}^{FJ\pm}$ basis and can be converted to the basis used in this paper using \refeq{eq: coordinatetransform}. As a shorthand we abreviate $\Yspp_{j,m} = y_{j,m}(\theta_0,\phi_0), \Sp_k = \sqrt{j+m+k}, \Tp_k = \sqrt{j-m+k}$.

\begin{flalign*}
&n_{jm}a_{jm0}^{\displaced,-}  = \frac{(-1)^m P r_0^{j-2}}{\viscratio}\sqrt{\frac{2j+1}{2j-3}} \bigg[\frac{(d_x+i d_y)^2}{4}\Tp_0\Tp_{-1}\Tp_{-2}\Tp_{-3}\Yspp_{j-2,-m-2}&&\\
&\hspace{30pt}-(d_x+i d_y) d_z \Tp_0\Tp_{-1}\Tp_{-2}\Sp_0\Yspp_{j-2,-m-1}-\frac{1-3d_z^2}{2}\Tp_0\Tp_{-1}\Sp_0\Sp_{-1}\Yspp_{j-2,-m}&&\\
&\hspace{30pt}+(d_x-i d_y) d_z\Tp_0\Sp_0\Sp_{-1}\Sp_{-2}\Yspp_{j-2,-m+1}+\frac{(d_x-i d_y)^2}{4}\Sp_0\Sp_{-1}\Sp_{-2}\Sp_{-3}\Yspp_{j-2,-m+2}\bigg]&&
\end{flalign*}

\begin{flalign*}
&n_{jm}a_{jm1}^{\displaced,-}  = \frac{(-1)^m P r_0^{j-1}}{2\viscratio}\sqrt{\frac{2j+1}{2j-1}} \bigg[\frac{(d_x+i d_y)^2}{2}\Sp_1\Tp_0\Tp_{-1}\Tp_{-2}\Yspp_{j-1,-m-2}&&\\
&\hspace{30pt}-(d_x+i d_y)d_z (j+2m+1)\Tp_0\Tp_{-1}\Yspp_{j-1,-m-1}-m(1-3d_z^2)\Sp_0\Tp_0\Yspp_{j-1,-m}&&\\
&\hspace{30pt}-(d_x-i d_y)d_z(j-2m+1)\Sp_0\Sp_{-1}\Yspp_{j-1,-m-1}-\frac{(d_x-id_y)^2}{2}\Tp_1\Sp_0\Sp_{-1}\Sp_{-2}\Yspp_{j-1,-m+2}
\bigg]&&
\end{flalign*}

\begin{flalign*}
&n_{jm}a_{jm2}^{\displaced,-}  = \frac{(-1)^m (2j+1)^{3/2} P r_0^j}{2j(2j-1)\viscratio}\bigg[\frac{(j+1)(2j+3)}{\sqrt{2j-3}}\bigg( \frac{(d_x+id_y)^2}{4}\Tp_0\Tp_{-1}\Tp_{-2}\Tp_{-3}\Yspp_{j-2,-m-2}&&\\
&\hspace{30pt}-(d_x+id_y)d_z\Sp_0\Tp_0\Tp_{-1}\Tp_{-2}\Yspp_{j-2,-m-1}-\frac{1-3d_z^2}{2}\Sp_0\Sp-{-1}\Tp_0\Tp_{-1}\Yspp_{j-2,-m}&&\\
&\hspace{30pt}+(d_x-id_y)d_z \Tp_0\Sp_0\Sp_{-1}\Sp_{-2}\Yspp_{j-2,-m+1}+\frac{(d_x-id_y)^2}{4}\Sp_0\Sp_{-1}\Sp_{-2}\Sp_{-3}\Yspp_{j-2,-m+2}\bigg)&&\\
&+\frac{1}{\sqrt{2j+1}}\bigg( -6\frac{(d_x+id_y)^2}{4}\Tp_0\Tp_{-1}\Sp_1\Sp_2\Yspp_{j,-m-2}+3(d_x+id_y)d_z (2m+1)\Tp_0\Sp_1\Yspp_{j,-m-1}&&\\
&\hspace{30pt}-2\frac{1-3d_z^2}{2}(j+j^2-3m^2)\Yspp_{j,-m}+3(d_x-id_y)d_z (2m-1)\Sp_0\Tp_1\Yspp_{j,-m+1}&&\\
&\hspace{30pt}-6\frac{(d_x-id_y)^2}{4} \Tp_1\Tp_2\Sp_0\Sp_{-1}\Yspp_{j,-m+2}\bigg)\bigg]&&
\end{flalign*}

\begin{flalign*}
&n_{jm}a_{jm0}^{\displaced,+} = \frac{(-1)^m P r_0^{-j-1}}{2(2j+3)\viscratio}\bigg[\frac{1}{(2j+1)(2j-1)j}\bigg(\frac{3(d_x+id_y)^2}{2}\Tp_0\Tp_{-1}\Sp_1\Sp_2\Yspp_{j,-m-2}&&\\
&\hspace{30pt}-3(d_x+id_y)d_z(2m+1)\Tp_0\Sp_1\Yspp_{j,-m-1}+(1-3d_z^2)(j+j^2-3m^2)\Yspp_{j,-m}&&\\
&\hspace{30pt}+3(d_x-id_y)^2\Sp_0\Tp_1\Yspp_{j,-m+1}+\frac{3(d_x-id_y)^2}{2}\Sp_0\Sp_{-1}\Tp_1\Tp_2\Yspp_{j,-m+2}\bigg)&&\\
&+\frac{1}{\sqrt{(2j+1)(2j+5)}}\bigg( -\frac{(d_x+id_y)^2}{4}\Sp_1\Sp_2\Sp_3\Sp_4\Yspp_{j+2,-m-2}&&\\
&\hspace{30pt}-(d_x+i d_y)d_z\Tp_1\Sp_1\Sp_2\Sp_3\Yspp_{j+2,-m-1}+\frac{1-3d_z^2}{2}\Sp_1\Sp_2\Tp_1\Tp_2\Yspp_{j+2,-m}&&\\
&\hspace{30pt}+(d_x-id_y)\Sp_1\Tp_1\Tp_2\Tp_3\Yspp_{j+2,-m+1}+\frac{d_x-id_y)^2}{4}\Tp_1\Tp_2\Tp_3\Tp_4\Yspp_{j+2,-m+2}\bigg]&&
\end{flalign*}

\begin{flalign*}
&n_{jm}a_{jm1}^{\displaced,+}  = \frac{(-1)^m P r_0^{-j-2}}{2 j (j+1)\sqrt{(2j+1)(2j+3)}\viscratio }\bigg[\frac{(d_x+id_y)^2}{2}\Tp_0\Sp_1\Sp_2\Sp_3\Yspp_{j+1,-m-2}&&\\
&\hspace{30pt}+(d_x+id_y)d_z(j-2m)\Sp_1\Sp_2\Yspp_{j+1,-m-1}+(1-3d_z^2)m\Sp_1\Tp_1\Yspp_{j+1,-m}&&\\
&\hspace{30pt}+(d_x-id_y)d_z (j+2m)\Tp_1\Tp_2\Yspp_{j+1,-m+2}\-\frac{(d_x-id_y)^2}{2}\Sp_0\Tp_1\Tp_2\Tp_3\Yspp_{j+1,-m+2}\bigg]&&
\end{flalign*}

\begin{flalign*}
&n_{jm}a_{jm2}^{\displaced,+} = \frac{(-1)^m Pj r_0^{-j-3}}{(j+1)(2j+1)^{3/2}(j+3)\sqrt{2j+5}\viscratio}\bigg[\frac{(d_x+id_y)^2}{4}\Sp_1\Sp_2\Sp_3\Sp_4\Yspp_{j+2,-m-2}&&\\
&\hspace{30pt}+(d_x+id_y)d_z\Tp_1\Sp_1\Sp_2\Sp_3\Yspp_{j+2.-m-1}-\frac{1-3d_z^2}{2}\Sp_1\Sp_2\Tp_1\Tp_2\Yspp_{j+2,-m}&&\\
&\hspace{30pt}-(d_x-id_y)^2 \Sp_1\Tp_1\Tp_2\Tp_3\Yspp_{j+2,-m+1}+\frac{(d_x-id_y)^2}{2}\Tp_1\Tp_2\Tp_3\Tp_3\Yspp_{j+2,-m+2}&&
\end{flalign*}

\bibliography{refsAD}

\end{document}